\documentclass[english,10pt]{article}

\usepackage[T1]{fontenc}
\usepackage[latin9]{inputenc}
\usepackage{graphicx}
\usepackage[table,xcdraw]{xcolor}
\usepackage{lscape}
\usepackage{babel}
\usepackage{hyperref}
\usepackage{url}
\usepackage{longtable}
\usepackage{varwidth}
\usepackage[center]{caption}
\usepackage[thinlines]{easytable}
\usepackage[acronym]{glossaries}
\usepackage[margin=1.3in]{geometry}
\usepackage[square,numbers]{natbib}
\usepackage[flushleft]{threeparttable}
\usepackage{array,booktabs,makecell}
\usepackage[font=small]{caption}
\usepackage{tabularx}
\usepackage{enumitem} 
\usepackage{amsmath} 
\usepackage{floatrow} 
\usepackage{titlesec}
\usepackage{arydshln}
\usepackage[noblocks]{authblk}
\usepackage{xcolor}
\hypersetup{
    colorlinks,
    linkcolor={blue!80!black},
    citecolor={blue!50!black},
    urlcolor={black!80!black}
    }
\usepackage{textcomp}
\usepackage{url}
\usepackage{authblk}
\usepackage{blindtext}
\usepackage{appendix}
\usepackage[colorinlistoftodos]{todonotes}
\usepackage{pdfpages}
\usepackage{amsmath,amssymb,amsfonts}
\usepackage{verbatim}

\newenvironment{myepigraph}
  {\par\hfill\itshape
   \begin{tabular}{@{}r@{\hspace{2em}}}}
  {\end{tabular}\par\medskip}

\newacronym{P2P}{P2P}{Peer-to-Peer}
\newacronym{TTP}{TTP}{Trusted Third Party}
\newacronym{IoT}{IoT}{Internet of Things}
\newacronym{DApp}{DApp}{Decentralized Application}
\newacronym{PoS}{PoS}{Proof-of-Stake}
\newacronym{PoW}{PoW}{Proof-of-Work}
\newacronym{NFT}{NFT}{Non-Fongible Token}
\newacronym{KYC}{KYC}{Know Your Customer}
\newacronym{CAGR}{CAGR}{Compound Annual Growth Rate}
\newacronym{DeFi}{DeFi}{Decentralized Finance}
\newacronym{DAO}{DAO}{Decentralized Autonomous Organization}
\newacronym{EVM}{EVM}{Ethereum Virtual Machine}
\newacronym{MVP}{MVP}{Minimum Viable Product}
\newacronym{LP}{LP}{Liquidity Pool}
\newacronym{Tokenomics}{Tokenomics}{Token Economy}
\newacronym{API}{API}{Application Programming Interface}

\begin{document}

\title{An Efficient and Decentralized

Blockchain-based Commercial Alternative

(Full Version)}

\date{October 15, 2022}

\author[1]{Marwan Zeggari\thanks{marwan.zeggari@lyzis.tech}}
\author[2]{Renaud Lambiotte \thanks{lambiotte@maths.ox.ac.uk}}
\author[3]{Aydin Abadi \thanks{aydin.abadi@ucl.ac.uk}}
\author[4]{Louise Axon \thanks{louise.axon@cs.ox.ac.uk}}
\author[5]{Mohamad Kassab
\thanks{muk36@psu.edu}}
\affil[1]{Lyzis Labs, USA}
\affil[2,4]{University of Oxford, UK}
\affil[3]{University College London, UK}
\affil[5]{Pennsylvania State University, USA}

\maketitle

\begin{abstract}
While online interactions and exchanges have grown exponentially over the past decade, most commercial infrastructures still operate through centralized protocols, and their success essentially depends on  trust between different economic actors. Digital advances such as blockchain technology has led to a massive wave of \textit{Decentralized Ledger Technology} (\textit{DLT}) initiatives, protocols and solutions. This advance makes it possible to implement trustless systems in the real world, which, combined with appropriate economic and participatory incentives, 
would foster the proper functioning and drive the adoption of a decentralized platform among different actors. This paper describes an alternative to current commercial structures and networks by introducing \textit{Lyzis Labs}, which is is an incentive-driven and democratic protocol designed to support a decentralized online marketplace, based on blockchain technology. The proposal, \textit{Lyzis Marketplace}, allows to connect two or more people in a decentralized and secure way without having to rely on a \textit{Trusted Third Party} (\textit{TTP}) in order to perform physical asset exchanges while mainly providing transparent and fully protected data storage. This approach can potentially lead to the creation of a permissionless, efficient, secure and transparent business environment where each user can gain purchasing and decision-making power by supporting the collective welfare while following their personal interests during their various interactions on the network.
\end{abstract}

 \clearpage

\thispagestyle{empty}
\small{\tableofcontents}
\thispagestyle{empty}

\normalsize{
\setcounter{page}{1}

\clearpage

\section{Introduction}

\begin{myepigraph}
Will not the Athenian people, too,
entrust their affairs to you \\
when they see that you have wisdom enough to manage them?\\[1.5ex]
Lysis, or Friendship\\
Plato
\end{myepigraph}

Trust is an essential element to ensure the proper functioning of Information Technology (IT) applications based on modern centralized systems.
In recent years, these IT applications have increasingly exposed some of the vulnerabilities and weaknesses inherently associated with centralized frameworks, primarily around single points of failure, multiple frauds and abuses, questionable (personal and/or financial) data handling, lack of transparency, and high fees.
These issues can ultimately lead to the creation of rules that users may not necessarily want to abide by \cite{yu2018iotchain}.
These flaws, as well as the sophistication of digital fraud, financial crimes and cyberattacks, are causing an increasing security risk and distrust of various market players, challenging the potential growth of many online-based businesses \cite{finfraud}.

Decentralization technologies can be seen as being, to some extent and through their various appealing features (e.g., a potential store of value, efficient transactional underlyings without any TTP control), a prospective long term solution to all the different issues related to centralized systems and all other types of structures where there may be a hierarchical ownership of user data. 
More recently, blockchain technology has gained an increasing attention from research and industry to implement several types of IT applications. 
This interest is mainly due to its core property that allows users to perform transactions without going through a TTP. In addition, it offers a transparent and fully protected data storage \cite{datastorage}.

Blockchain technology was introduced by Stuart Haber and W. Scott Stornetta in 1991 \cite{haber1990time}, but its first implementation appeared during the year 2008 when Satoshi Nakamoto proposed Bitcoin as a decentralized Peer-to-Peer (P2P) cryptocurrency application \cite{nakamoto2008bitcoin}.
Bitcoin has undoubtly made blockchain technology an important trend in decentralization and a point of change for understanding modern P2P interactions \cite{zhang2019security}. 
The introduction of this disruptive technology of storage and transmission of information has then allowed the creation and advancement of initiatives in many fields (e.g., finance, logistics, art, real estate, IoT, and identify management) that are an integral part of our modern society \cite{realworldblockchain}, as well as the advent of more autonomous 2.0 applications without any hierarchical system as such: distributed/decentralized applications (dApps). 
These dApps operate independently and autonomously without any TTP, for instance thanks to the use of smart contracts that run on a decentralized and distributed computing system, the blockchain \cite{dApps}. 
DApps leverage all the strengths of blockchains (i.e., trust, transparency, visibility, traceability, immutability, resilience, privacy improved, and intermediary costs suppressed). Their deployment, with appropriate economic incentives and technical bases, gives each individual the potential to independently contribute to the common welfare, while essentially pursuing their own economic and decision-making/governance concerns within the system.

\subsection{Expectations, and Contrast to Existing Commercial Approaches}

Lyzis's core objective is to build a stable and fair business environment surrounding a dApp, based on incentives and an optimal balance between participation and value extraction.
The main focus is to relieve all sellers and buyers - and more generally, e-commerce actors around the world - of the need to trust both their interlocutor and the platform supporting the operation and ensuring its smooth running.
This paradigm shift allows for a commercial system that ensures freedom and economic stability for all participants.
Lyzis Marketplace, a blockchain-based platform where any user can buy and sell physical assets according to predefined conditions using smart contracts, is mainly designed to solve the trust issues faced on centralized e-commerce systems.

To achieve this, the blockchain's scalability, security, transparency, and decentralization features are leveraged, using Proof-of-Stake (PoS) as a consensus algorithm on a permissionless blockchain. 
This is essentially how our proposal differs from conventional platforms and centralized shopping structures.
Also, to differentiate itself technically and strategically from the existing blockchain industry players, our solution supports \textit{(i)} the integration of a trust system within Lyzis Marketplace that operates as an underlying decentralized service providing a feedback mechanism for end-users (main competitors' weakness), \textit{(ii)} the use of appropriate economic incentives, \textit{(iii)} the application of a set of game-theoretic mechanisms, \textit{(iv)} the adoption of an adequate initial go-to-market strategy, and \textit{(v)} the possibility of user involvement in the platform's decision-making process.

As an additional benefit, Lyzis Marketplace would be an entry point to mainstream users, giving them a first, secure experience with the  web3 and its ecosystem.
Overall, it is our belief that Lyzis provides the necessary technological infrastructure, and economic/underlying incentives, required to give additional purchasing and decision-making power to both buyers and sellers, and would allow to alleviate the critical limitations of today's trust-based, centralized marketplaces.

\section{Contextualisation}

This section first provides in \ref{background} a brief introduction to blockchain technology, its evolution and technological changes as well as in \ref{blockusecases} its various applications in many fields. It also discusses in \ref{intromarket} a parallel between this disruptive technology and marketplaces such as Lyzis Marketplace for the exchange of physical assets. The remaining section then outlines the several challenges surrounding decentralized marketplaces in Section \ref{marketchallenges} and finally provides a summary of the proposal in Section \ref{overviewanddata} by discussing the various solutions\footnote{The solutions are based on the flaws found within centralized systems, as identified by the recently collected internal market data (see Appendix \ref{AppendixA1} for more details on the internal market survey).} brought by the project.

\subsection{Background}\label{background}

Although technologies embedded in blockchain such as Merkle trees (Binary and Patricia) \cite{haber1990time} and cryptographically secure chains \cite{cryptochains} were developed in the 90s \cite{TrustworthySmartSystems}, the first concept of blockchain as such was introduced by S. Nakamoto in 2008 \cite{nakamoto2008bitcoin}.
It was proposed to support the functioning of a P2P version of digital cash operating without third party control.
It came as a profound technological and libertarian response to the crisis of 2007-08, also known as the subprime mortgage crisis, and to the traditional banking and financial system as a whole \cite{tradbank}, leading to a distrust in any (financial) authority. Later on, to improve the functionality of Bitcoin, a new generic cryptocurrency platform, called Ethereum \cite{ETHpaper} was proposed. 

Cryptocurrencies, such as Bitcoin and Ethereum, beyond offering a decentralized currency, support computations on transactions.
In this setting, often a certain computation logic is encoded in a computer program, called a \textit{smart contract}. 
Although Bitcoin, the first decentralized cryptocurrency, supports smart contracts, the functionality of Bitcoin's smart contracts is very limited, due to the use of the underlying programming language that does not support arbitrary tasks \cite{paymentResolution}.
To overcome this limitation, Ethereum, as a generic smart contract platform, was designed. Thus far, Ethereum has been the most predominant cryptocurrency framework that lets users define arbitrary smart contracts. This framework allows users to create an account with a unique account number or address.
Such users are often called external account holders, which can send (or deploy) their contracts to the framework's blockchain. In this framework, a contract's code and its related data are held by every node in the blockchain's network.
Ethereum smart contracts are often written in a high-level Turing-complete programming language called \textit{Solidity}. The program execution's correctness is guaranteed by the security of the underlying blockchain components. To prevent a denial of service attack, the framework requires a transaction creator to pay a fee, called \textit{gas}.
To address Bitcoin's consensus limitations, i.e., Proof-of-Work (PoW), such as inadequate scalability and energy expenditure, Proof-of-Stake (PoS) was first proposed\footnote{Also originally developed by S. King and S. Nadal in 2012 in the Peercoin blockchain \cite{PPcoin}, in a hybrid scheme with PoW.} as a standalone protocol in the Nxt blockchain in 2013 \cite{Nxt}.
PoS  replaces the competition of PoW by randomly selecting the stakeholders to append to the blockchain and has thus seen an exponential growth since its inception \cite{survey}.

There are also, mostly, two types of blockchains: permissioned and permissionless \cite{TrustworthySmartSystems}. 
In permissioned blockchains, it is a designated authority that is in charge of pre-approving all the network participants, making it open to a very limited number of users.
As this approach is limited to specific applications (e.g., internal business management) and as it does not directly concern the Lyzis Marketplace, it will not be considered in this paper and we will only focus on permissionless blockchains.
In a  permissionless model, the network is  open to all participants and some must reach a consensus to make a decision \cite{permissionandpermissonless}.
Together, the many initiatives that have built on the permissionless blockchain and with the use of PoS as a consensus algorithm have shown the significant advantages that can be gained for attributable governance, data security, privacy together with a significant reduction of intermediary costs for the realization of an exchange, contract, tracking, and more. This may ultimately solve the problem of trust traditionally expressed by users towards centralized applications, as well as the inherent lack of transparency of their rules and conditions, by allowing all users to be hierarchically equal and to rely entirely on blockchain-based protocols.

\subsection{Blockchain Applications}\label{blockusecases}

The blockchain technology has a wide range of possible use cases, and can be applied in all sectors and scenarios where the involvement of a TTP is not desirable (and a P2P system is needed) to manage transactions, exchanges, pre-defined terms, and more, with characteristics such as transparency, integrity, immutability, security and confidentiality \cite{TrustworthySmartSystems}.
Here are some possible areas of blockchain application that have successfully emerged since bitcoin's initial implementation \cite{nakamoto2008bitcoin}.

\textit{Financial Contracts}: The rules governing any type of financial product or service can be encoded in the form of a smart contract \cite{fincontr} to facilitate claims settlement and automated financial transactions. DeFi applications can for example address a very large market of financial transactions without the need for banks, which can be seen as more advantageous than traditional finance mainly due to the openness and accessibility of all types of actors, and the (24 hours) continuous and without-border availability.

\textit{IoT Monitoring}: Blockchain technology can enable the digitization of assets using IoT sensors so that organizations of all kinds can label their assets and provide a more transparent tracking system \cite{iotmonit}. Digitization then identifies the location and status of any item and the blockchain can therefore store, manage, protect and transfer all this data internally in a transparent, accessible and immutable way.

\textit{Public Sector}: It is also possible to apply blockchain technology to the public sector, e.g., by using its transparency properties to increase trust in voting processes \cite{votingprocess}. Here, participants can participate and vote in a decentralized system without a single point of weakness in elections of any kind, then significantly improving the various transparency deficiencies known until now \cite{trustissuevote}.

\textit{Health Care}: Blockchains can  verify, secure and share useful data, e.g., following the Covid 19 crisis to confirm the identity, vaccination status and permission to leave the home of a person affected by restrictive measures \cite{ChamolaHealth,IBM} or to prevent healthcare worker fatigue and promote targeted early intervention \cite{Dhillon}. This technology is  an ideal tool for secure tracking and tracing, enabling on-demand design and delivery of personalized medical equipment \cite{trackmanagehealth} and identification of potential patients or contagion niche(s) with transparent sharing of this information to public health authorities, hospitals, etc. \cite{HealthCare}.

As previously mentioned, the potential application cases for blockchain technology are countless and its usefulness points to a wide range of possibilities. In the next section, we will look more in detail at its applicability to decentralized marketplaces, which is at the core of the Lyzis proposal.

\subsection{Decentralized Marketplaces: Value Insight}\label{intromarket}

Web insecurities, questionable data processing (banking and personal), unauthorized withdrawals, the need to audit and/or trust a transaction partner, among others, lead to critically define the limits of centralized marketplaces.
Decentralized marketplaces that connect users in a P2P manner are then exactly what blockchains are all about. Thanks to the use of smart contracts, sellers cannot be discriminated against by the system, as the rules are fixed, known to all and unchangeable by the smart contract. Importantly, the rules of the system, in this case the Lyzis Marketplace, are publicly verifiable by each individual wanting to be involved in an exchange. Moreover, the transaction data are available on the blockchain through the registry and is thus open to all participants, which also makes a potential audit procedure much simpler for the platform. 
As an additional advantage, a cost analysis performed on decentralized marketplaces  \cite{decenmarket} has shown that it is not expensive for a seller to list an item and sell it, which lowers prices for buyers even more with the use of side-chains (e.g., Polygon Chain) and PoS as a consensus algorithm, reducing the overall costs of instantiating a smart contract. 
\\
Moreover, Ethereum being a reference as a framework for the construction of protocols, the latter ensures the security of users' data, guaranteeing that no user can acquire the information of another user and thus identifies each actor only by his public key. Within the framework of a decentralized object exchange, only the party concerned by the sending of the physical object (the seller) has at a given moment the address of his correspondent (the buyer) in order to send him the package, but no other actor can obtain any other information/data concerning the various participants of the marketplace, except for the public address and the transactions related to it. 
Critically, all exchanges are done only using digital assets (i.e., tokens or stable assets). Through the use of these elements underlying the blockchain, none of the actors (buyers or sellers) needs to allocate any level of trust towards their transaction partner and the platform on which they exchange. The environment becomes totally self-executing and most importantly, asymptotically trustless.

Along these considerations, a decentralized marketplace as proposed in this paper is not subject to the various weaknesses of systems that may depend on specific actors, which may lead to reduced access to resources, breakdowns, lack of incentive mechanisms to prevent corruption or inefficiencies, or any other type of vulnerabilities possibly encountered in traditional centralized marketplaces (e.g., Facebook Marketplace, Ebay, Craiglist). 
Another advantage is scalability\footnote{Nodes in the distributed system are connected to each other and can conveniently share data. Additional nodes may easily be added to the distributed system and hence scaled up as needed.}, which is essential as the volumes of assets exchanged across hierarchical structures rapidly increase. In this instance, distributed architectures without central control are more effective than conventional centralized systems.
Note that relying on blockchain technology does not necessarily make a marketplace decentralized. For instance, we have recently seen in the web3 industry the widespread appearance of NFT marketplaces, with notably Opensea \cite{Opensea} as a reference when it comes to trading non-fungible assets, but it is again a centralized service with a single point of failure and several applications rely on it. It is then the source of a systemic risk, if an application goes down, it drags the others with it.
The main aim of Lyzis' marketplace, as a decentralized platform, is to set the standard for P2P physical asset exchanges, such as Opensea for NFTs, but ensuring a decentralized operation in terms of exchanges, thus staying in line with the values of the blockchain ecosystem while allowing people to gain access and benefit from decentralization.

\subsubsection{Various Challenges}\label{marketchallenges}
 
 In this section, we give an overview of the main challenges that may hinder the successful realisation of a decentralized marketplace \cite{challengesdemarket}, which includes technical, structural and strategic challenges.

\begin{itemize}
\item \textit{Security} - or rather how to establish a good trust system while dealing with: \textit{(i)} cybersecurity issues and \textit{(ii)} strategic users trying to cheat the system and/or manipulate the market. This is  the first challenge to consider.  The implementation of several mechanisms, including economic and organizational incentives are required (e.g., reward system, reputation within the network), as well as a thorough analysis of the behavioral scope of each actor involved within the platform. This will help by preventing negative actions of agents while maximizing the contribution of each to the P2P network, assuming we study all possible attack parameters \cite{decenmarket}, such sybil attacks \cite{SybAtt1}. They are most important and most common attacks on reputation systems; they consist in having a malicious node behaving as if it were a larger number of nodes, i.e., by impersonating other nodes or simply by claiming false identities \cite{SybAtt2}. The attacker can then generate an arbitrary number of additional node identities within the network with the use of a single physical device \cite{SybAtt3}; to compromise the reputation system established within the network and gain the majority of influence over it, by performing illegal actions (in relation to the rules and laws established in the network). This can be prevented with a good matching engine and general cybersecurity mechanisms to ensure a good operation.
\item \textit{Adoption} - It is also a challenge to consider \cite{adoptionblockchain}. While platforms like Ethereum and other blockchain projects have been adopted by companies, by a wide range of individuals for their various uses, as well as by some institutions, these technologies are not yet mainstream and remain mysterious for the common people. It is thus critical to implement many features to facilitate adoption among several categories of users (e.g., custodial and non-custodial wallets and web 2 like interfaces) in order to facilitate the access to the platform while staying in line with more classical web interfaces. It is also preferable for a P2P structure to first trigger adoption and engagement by a specific niche of users, e.g., by reducing the categories of products available in order to avoid addressing too broad a user base in the first phase, and thus watering down interest and value. The acceptance of digital assets among mainstream internet users as well as the growth of many well-targeted initiatives are expected to play a critical role in adoption and hence to help overcome this challenge over time.
\item \textit{Content Control} - Decentralized marketplaces have, by nature, no real central authority to operate and prevent the downloading of content, services and items for sale that may be, e.g., dangerous, illegal, criminal. It is therefore necessary to implement features to control the type of ads that can be posted on it - these measures being mainly aimed at preventing the use of the marketplace in question as a black market and to nuance/dissociate it from the similarities with some darkweb platforms, i.e., payment in digital assets, anonymity (visibility of the public address only), exchanges of all kinds, and more. Some of these measures may include, among others, the implementation of a KYC (Know Your Customer\footnote{Know Your Customer (KYC) guidelines in financial services require professionals to make efforts to verify the identity, suitability and risks of maintaining a business relationship.}), the need to activate a wallet through staking before carrying out exchanges to increase the seriousness of the marketplace actors, the need to reach a certain consensus on the available categories and/or the posted ads, and so on.
\item \textit{Vendor Attraction} - Sellers and their involvement are key pillars of a well-functioning marketplace to ensure \textit{(i)} available content of a certain quality/standard and \textit{(ii)} a secure and distributed environment that does not require any level of trust from the buying user's perspective.
The  attraction of high-quality vendors is  a challenge that can be remedied by the implementation of numerous incentives (or repellents) pushing the sellers to adopt a correct, honest or even beneficial behavior for the ecosystem. It is essential to take into consideration their personal interests and their possible fields of action within the environment in order to maximize their implications and their beneficial impacts on the platform, and thus obtain a healthy marketplace (e.g., by rewarding the deposit of an ad in the first phase in order to increase the offer and thus drive the demand upwards, by imposing a minimum price threshold for a deposited item in order to guarantee their interests and not be overwhelmed by useless and unsaleable items).
\item \textit{Optimal Balance} - In parallel to \textit{Vendor Attraction},  it is also necessary to reach within the marketplace a stable equilibrium, ideally growing on the medium/long term, between the deposit of ads, i.e., the inflow of sellers, and the number of purchases made, i.e., the inflow of buyers. This balance  ensures a fairly constant daily flow of exchanges and interactions, and is necessary to make the project viable and attractive. Achieving this goal is still, as of today, a considerable challenge for blockchain-based marketplaces, partly due to various factors such as the interest and value of the items deposited on the platform, the ever-changing legal environment around projects (presence restrictions), a misdirected go-to-market strategy and an overly large field of targeted users, the volatility of cryptocurrencies, etc.
\item \textit{Secure Communications} - The communication models established within decentralized environments record and transmit each information exchanged and interaction performed to the blockchain in a public way. In the case of a decentralized marketplace, only brief interactions are necessary between two different parties to, among other things, agree on the terms of trade and to keep track of a follow-up and/or settle a dispute, always within specific pre-defined timeframes. In order to avoid any inefficiency and to facilitate private exchanges between individuals, it is preferable for creators to implement only a secured instant messaging (IM) service, e.g., in order to solve the challenge of offering a secure environment while managing more simply certain cases of dispute in internal arbitration.
\end{itemize}

\subsection{Findings}\label{overviewanddata}

Our vision for Lyzis Labs is supported by the documented flaws and recurrent cases of fraud that can be encountered/experienced on traditional hierarchical and centralized commercial protocols. These inherent limitations give ground for the implementation of a P2P trustless exchange model and of a system based on decentralization, using blockchain technology to operate. The remainder of this section summarizes our key findings, and additional information about the market can be found in Appendix \ref{AppendixA1}. 

\subsubsection{Factual Market Observations}

The Covid-19 pandemic has dramatically and systemically affected  our  society, impacting companies, individuals and institutions simultaneously. It has also reinforced the essential role played by the global internet for the continuous functioning of the world economy, and in our daily life more generally.
However, the global digitalization of the economy and the significant increase in the use of mobiles, applications, dematerialized services, and so on raise several crucial IT security issues (e.g. data privacy, transaction security, company transparency). 
A majority of these issues can naturally be solved by adopting a model based on blockchain technology, leading to increased demand from various companies to adopt decentralized ledgers for their specificities.

\vspace{+0.3cm}

\textit{The use of the internet when interacting (transactionally or otherwise) is vital nowadays to keep the economy going and more and more players are finding their way there - by necessity and/or not.} 
\vspace{+0.3cm}
\\
We are seeing significant growth at all scales of online transactions and in the case of online commerce, the global e-commerce sales market totals \$5.5T (2021-2022) - of which retail sales represent 21\% of the share - and these numbers are expected to grow exponentially over the following years with forecasts of reaching a total of \$7.5T by 2025 for the global e-commerce sales market, with 24\% for the retail sales share \cite{statistaecomm1, statistaecomm2}. The upcoming global online retail landscape represents an ideal opportunity for Lyzis considering the significant sizes of the target markets.

\vspace{+0.3cm}

\textit{Several trust issues are inevitably and systematically associated with centralized IT protocols of all types.}
\vspace{+0.3cm}
\\
On the current centralized/hierarchical protocols dedicated to commercial exchanges of all types, many flaws and vulnerabilities are well identified, the most common being:
    \begin{itemize}
    \item[--] Cybersecurity (i.e., CB hacking, CB registered without authorization, debit without reason/authorization, locked funds, effective debit but purchase order not accounted for, use of personal information including phone number (transmission), global non-transparency and lack of user control over their data and network rules, cancellation/deletion of submitted reviews - unreliable rating system (faked/bought), misleading item descriptions, unjustified penalties applied, identity verification problem and consequent lack of access, bank account number and/or social number to be provided);
    \vspace{-0.15cm}
    \item[--] Inefficient/deplorable after-sales service (i.e., ignoring complaints, unacceptable delays, no protection for sellers but only for buyers, defective dispute management, non-management when there is no response from the seller, no study in dispute/automatic responses);
    \vspace{-0.15cm}
    \item[--] Abuse of guarantees (i.e., either from third parties like PayPal or directly from the relevant platform);
    \vspace{-0.15cm}
    \item[--] Defective, damaged or non-conforming receipt and/or impossible return;
    \vspace{-0.15cm}
    \item[--] Ineffective refunds (i.e., none and/or only a small portion);
    \vspace{-0.15cm}
    \item[--] Ineffective order cancellation;
    \vspace{-0.15cm}
    \item[--] Package noted delivered but never received;
    \vspace{-0.15cm}
    \item[--] Possible physical meetings (e.g., assaults, thefts);
    \vspace{-0.15cm}
    \item[--] Users' account desactivated without any reason and abusive procedures;
    \vspace{-0.15cm}
    \item[--] Ads automatically not approved (i.e., without valid reasons).
    \end{itemize}

The \textit{"Trust Issues"} notion evoked along this paper refers then directly to the flaws and vulnerabilities listed above.

\vspace{+0.3cm}

\textit{There is currently a significant move of companies towards blockchain solutions in order to remedy trust issues.}
\vspace{+0.3cm}
\\
The rapid adoption of decentralized solutions is essentially driven by effective operational underpinnings to alleviate various known problems (e.g., TTP-free exchange), by investment/speculation vehicle and by the emerging application/use of decentralization in various domains (i.e., travel, healthcare, finance - DeFi, and education).  This has led the blockchain market to account for \$4.67 billion in 2021 and is expected to grow to \$163.83 billion by 2029 with a compound annual growth rate (CAGR) of 85.9\% from 2022 to 2030 and revenue forecast in 2030 of \$1,431.54 billion \cite{report}. The world is taking the path of decentralization and every day new players are finding their way into this market at all levels, hence allowing Lyzis to draw the necessary attraction and establish itself in this rapidly evolving market.

\subsubsection{Lyzis Labs' Related Solutions}
If we break down and classify - for reasons of concreteness - the flaws and vulnerabilities faced in centralized marketplaces, we get the next 3 main issues that are closely correlated: \textit{Trust}, \textit{Control}, and \textit{Security/Safety}.
Tab.\ref{Tab0} shows the different categories and the solutions provided by Lyzis.

\begin{table}[!h]
\resizebox{\textwidth}{!}{%
\begin{tabular}{|l|l|l|}
\hline
 & Issue & Lyzis' solution \\ \hline
\textit{Trust} & \begin{tabular}[c]{@{}l@{}}A buyer/seller needs to verify and ensure \\ the credibility of both a seller/buyer and \\ the intermediary platform to engage in an \\ interaction - there is an ultimate need for \\ trust in all parties involved.\end{tabular} & \begin{tabular}[c]{@{}l@{}}We use blockchain technology and hence \\ smart contracts as the operative basis \\ to provide users with a secure, \\ autonomous and trustless interaction \\ without any TTP intervention.\end{tabular} \\ \hline
\textit{Control} & \begin{tabular}[c]{@{}l@{}}A centralized intermediary platform\\ implements rules in a non-transparent way\\ that users may not know of or adhere to \\ without any possibility of expression and/ \\
or intervention.\end{tabular} & \begin{tabular}[c]{@{}l@{}}We ensure transparency by design, using \\ smart-contracts, and we provide gover-\\nance mechanisms to align the values \\ of the system with those of the users.\end{tabular} \\ \hline
\textit{Security/Safety} & \begin{tabular}[c]{@{}l@{}}The use of third parties, unauthorized use of \\ bank/financial data, fraudulent users - among \\ other things - lead to an unsafe environment \\ for a user wishing to perform online \\ commercial transactions.\end{tabular} & \begin{tabular}[c]{@{}l@{}}The use of wallets during interactions as \\ well as the implementation of several trust \\ mechanisms drawn from game theory  \\ provide users with an ecosystem \\ where everyone can contribute to the \\ common welfare by promoting their \\ respective concerns while feeling safe.\end{tabular} \\ \hline
\end{tabular}%
}
\caption{Key issues with centralized platforms and Lyzis' solutions.}
\label{Tab0}
\end{table}

\section{Lyzis Network}

The Lyzis network is an implemented system whose functionalities are built on technological and economic paradigms that take into account agents' behavior according to their preferences, payoffs and rationality (equilibrium analysis). This framework makes it possible to build a system where the adoption of the central platform is linked to the needs and diverse participation desires of a large range of actors.
This makes it possible to drive and maximize the main function of Lyzis Marketplace, which is the exchange of physical objects between individuals in a decentralized and P2P way benefiting from the benefits of blockchain technology and smart contracts.

The remainder of this section is organized as follows. Section \ref{over} provides an in-depth overview of the Lyzis network and Section \ref{feat} introduces the various built-in features for the future evolution.

\subsection{Overview}\label{over}

Our vision for Lyzis Labs is based on two well-defined core properties, a permissionless environment suitable for any user type and the ability to solve specific transactional issues plaguing today's business platforms. 
We believe that these two elements are essential for the continuous increase of entrants into the Lyzis ecosystem, by making it a safe 
entry point for a user into a decentralized environment (i.e., an exchange platform) and a stimulating one to ensure engagement via rewards, benefits, and personal advantages.
Lyzis Marketplace allows two or more people to be connected in a P2P, decentralized and secure way without having to go through a TTP to mediate physical object exchanges. It provides transparent and fully protected data storage thanks to the use of the blockchain technology. By implementing this platform, we therefore want to solve the major trust issues involved in all online commercial interactions and/or structures, being a key element ensuring the sustainability of transactions. 

Indeed, the trust problem in e-commerce is much more critical than in traditional commerce transactions due to a greater uncertainty, which results in  more and more serious frauds and  anxiety on the part of consumers \cite{trustecommerce, trustdistrust}. 
Several factors can explain this  anxiety, notably the fact that in the context of  e-commerce platforms (centralized or decentralized), the majority of purchases are made without face-to-face interactions and often at rather large geographical distances - buyers therefore do not have the possibility of touching the product directly and testing it in person \cite{impulsepurchasing, Shopattitude}. 
With the increasing popularity of e-commerce, several studies have  explored the determinants of an online buyer's trust in online sellers \cite{drivertrust}. 
For instance, some of the e-commerce literature focuses on buyer-related elements such as shopping experience \cite{shopexp}, hedonic value \cite{hedonic} and trustworthiness \cite{trustworthiness}. Many more have also focused primarily on website-related elements such as security and privacy \cite{websiterelated}, website quality \cite{websitequality} and perceived usefulness \cite{perceivedusefulness}.
This body of work leads us to conclude that the essential factors/drivers for a well-functioning marketplace, with or without authority, are mainly based on the willingness to trust the different actors, a problem that has not been satisfyingly resolved so far due to the complexity and broad nature of online trust \cite{drivertrust}.

Lyzis Marketplace is built on the principle \cite{cognitive} that the online shopper's willingness to trust one another is essential for its success. 
Building on \textit{postulates} from \cite{drivertrust}, we consider that within the Lyzis Marketplace and more globally the Lyzis ecosystem:
\begin{itemize}

\item\textit{Online trust\footnote{We define \textit{online trust} as \textit{the extent to which an online consumer will accept the vulnerability and trustworthiness of relying on the credibility, honesty, and competence of an online seller to deliver in the future}. When using a P2P structure, we hypothesize that the online consumer will rely instead on the principles of decentralization and autonomy, as well as the trustworthiness provided by these, rather than the honesty and competence of individual peers.} will have a  positive effect on an individual's attitude\footnote{As \cite{Attitude} outlined, \textit{attitude} refers here to the \textit{immediate favorable or unfavorable evaluation of a particular behavior}.}.} According to \cite{PurchBehav}, the perceived risk associated with online sellers on e-commerce platforms is the degree to which an individual believes in potential losses when purchasing a product and/or service on a particular e-commerce site (e.g., delivery of a defective product and unauthorized sharing of personal information). The application and use of a smart contract to manage the exchanges of an e-commerce platform e.g., Lyzis Marketplace here, significantly reduces the possible degree of loss to which any seller may be exposed. The resulting increase in online trust generally translates into a positive effect on the attitude of sellers and makes them more inclined to complete a transaction. Also, e-commerce research shows that as the risks associated with e-commerce decrease, attitudes toward Internet shopping become more positive. 

\item\textit{Online trust will have a negative effect on risk perception.} Alongside the positive effect that online trust can have on the attitude of an individual, online trust also has a subsequent negative effect on the perception of risk within the network. Indeed, numerous studies regarding e-commerce adoption show that trust is a major determinant of online shopping attitudes \cite{websitequality, Privacy, consumbehavgroc}. For this reason, we consider that perceived trust during interactions on an online commerce exchange network, specifically purchases, exerts a positive influence on consumer attitude formation \cite{consumbehavgroc}. As such, the risks perceived by the different actors (sellers/buyers) depend entirely on the vision and the trust given by them towards the structure of Lyzis Marketplace and its functioning.
    
\item\textit{Risk perception will have a negative effect on an individual's attitude.} According to the theory of planned behavior, behavioral intention relates to the individual's subjective probability of performing a specified future behavior under conditions of one's volitional control \cite{plannedbehav}. Different studies put forward the prospect of a potential negative association between a buyer's perceived risk and his purchase intention towards a platform \cite{privacyimpact, producteval, agriprodcons}: indeed, the perceived risk of a consumer towards a seller, and even more globally towards a platform, considerably influences the potential purchase intention of the consumer. For this reason, a clear definition of the potential actions for each actor  (buyers/sellers), their perceived risks, and the planning of their behaviors within an exchange environment are a prerequisite for a proper transactional and participative functioning.

\item\textit{Individual's attitude will have a  positive effect on purchase intention.} Numerous studies have pointed to the fact that trust significantly and positively affects the purchase intentions that can be expressed towards online sellers by buyers of all types \cite{antec, custominfo, advertperspect} and that the perception of trust that can be attributed to a platform is a key element that predicts online purchase intentions \cite{bloggereffects}. This observation suggests that an attitude expressed in a beneficial way by a user, i.e., the adoption of an honest behavior, and a reward to this same user for such attitude, contributes strongly to the positive impact that can be observed on his purchase intentions - if a user (seller or buyer) is induced to choose a beneficial and honest attitude, then its interlocutor (seller or buyer) possesses a crucial piece of information: the adoption of beneficial and honest strategies contribute to the personal interests of both parties.

\item\textit{Perceived reputation\footnote{Here, we assume that \textit{perceivable reputation} is defined as a consumer belief that allows an online seller (trustee) to be subjectively perceived as trustworthy by an online buyer (trustor).} will have a positive effect on online trust.} Since trust in a seller and their reliability is essential to reducing the risk that a buyer may perceive individually when shopping and interacting on a platform, the reputation of the seller(s) is a key determining a buyer's shopping intentions. Indeed, higher levels of e-shopper trust lead to lower levels of perceived risk when shopping online \cite{onlinegrocershop}. Consequently, reputation being a strong point and an essential element in guaranteeing a certain level of trust expressed by a buyer towards a seller and then more globally towards a platform, several mechanisms in this sense, i.e., rating system, deposit of reviews, are beneficial and proposed within the Lyzis Marketplace.
    
\item\textit{Online trust will have a positive effect on purchase intention.} Taking the previous factors together, the trust in the Lyzis Marketplace and in the different actors (i.e., sellers and buyers) has  a significant, important and positive impact on the buying intentions of the buyers towards the sellers. In a decentralized model with a built-in trust system, as in the Lyzis Marketplace, buyers will have no reluctance and will thus be much more inclined to interact on the network. This intention to buy is anticipated to result in increased operations and eventually in the achievement of an equilibrium point representing a level of transactional viability within the platform.
   
\end{itemize}

\subsection{Environment Features}\label{feat}

Lyzis Labs support the development and implementation of various technologies and features within its protocol that are critical to network integrity, access and performance. Three essential elements/parameters must be guaranteed in order to achieve a decentralized, scalable, secure, and immutable transactional, financial and commercial network: \textit{utility}, \textit{participation} and \textit{stability}.

\begin{itemize}
   \item \textbf{Utility:} To reach a growing and stable usage volume in the long term while having the necessary resources to do so, a decentralized project such as Lyzis Labs built on top of the blockchain technology must be able to provide specific and adapted use cases other than a simple fundraising involvement, and then capture the interest of a large field of users by bringing added value.  Since utility contributes to the intrinsic value of each of the elements created within an environment, we will introduce a native utility token (LZS) primarily designed to function as a medium of exchange, thus providing the incentives, features and necessary use cases (e.g., buying and selling on the marketplace, staking, future fee reduction) needed to ensure the provision of sufficient resources for healthy growth and the development of a structured environment thereafter (see Section \ref{LZS}).

  \item  \textbf{Participation:} It is also necessary to increasingly solicit the participation of all participants to ensure a successful  development.  As the degrees and qualities of solicitation/involvement vary greatly, several aspects are to be taken into account: informing, consulting (e.g., feedbacks, pools), soliciting (e.g., voting), collaborating (e.g., joining in the expansion and actively participating in it), and power delegation/empowerment (the gaining of decision-making power in order to best influence the future of the ecosystem). In this model, 
    the participants can and must  actively build and shape the environment, participating at their own scale in the different levels of involvement in order to accentuate and extend the central decision tree of the ecosystem. Hence the LZSP - a governance and participation token designed to foster good functioning and reward different agents with decision-making power - is  introduced to ensure a  balance in the various decisions as well as to incentivize favorable behaviors to reduce inefficiencies (see Section \ref{LZSP}).
    
    \vspace{+0.05cm}
    
  \item  \textbf{Stability:} Independently and in addition to the two previous elements of utility and participation, it is also necessary to bring within the ecosystem a certain monetary stability allowing essentially to reduce the perception of the risk to which  new entrants (sellers and buyers) are exposed when setting up crypto-currency accounts, and to allow  equal exchanges of values between all types of actors according to their wills and preferences. To palliate the well-known volatility of blockchain currencies, Lyzis will incorporate a stablecoin (LZDC), to mainly enable payments of a non-fluctuating nature between users interacting on the network and to allow simplified conversion to more conventional currencies (see Section \ref{LZDC}).
    \end{itemize}
    
These main factors together will contribute  significantly to the successful development of the marketplace, to its adoption by different types of users and to its sustainability by bringing confidence and ensuring healthy growth.
In addition, the Lyzis technological system being decentralized at its core, it will swiftly allow the deployment of:

\begin{itemize}
    \item[$-$] \textit{Lyzis Chain}: A PoS Sidechain\footnote{A sidechain is a separate blockchain which runs in parallel to the main Ethereum network and operates independently. It is linked to Mainnet by a two-way bridge.} connected to Ethereum Mainnet (\textit{EVM}\footnote{The Ethereum Virtual Machine (EVM) is a virtual component within each Ethereum node able to run bytecode for contracts.} Compatible).
    
    \item[$-$] \textit{Lyzis DAO}: On-chain Voting, Multi-sig Wallet (Committee) and open to LZSP holders.
    
     \item[$-$] \textit{Lyzis Dashboard}: DAO\footnote{A Decentralized Autonomous Organization (DAO) is a collectively-owned, blockchain-governed organization working towards a shared mission \cite{ethdao} (see Section \ref{DAO}).} Access, Wallet Connection\footnote{Users can import a wallet (non-custodial, e.g., Metamask) or use a Lyzis wallet (custodial) (see Section \ref{custodandno})} and other project related specifics.
\end{itemize}

\section{Lyzis Martketplace}

By using the blockchain's security, decentralization and scalability features and using PoS as consensus algorithm, Lyzis Marketplace, a blockchain-based platform, aims to overcome the trust issues faced by traditional centralized e-commerce protocols.

By design and compared to similar but centralized systems, Lyzis' marketplace is able to offer several advantages to all market participants, including trust and hence security, confidentiality, transactional integrity and significantly reduced transaction costs.
The decentralization and the use of a P2P model disrupt the paradigm of today's traditional/centralized marketplaces in which a large intermediary company controls every aspect of the exchange, from product listings to price discovery, product search, logistics, and the customer experience.
Several participants of all categories - web3 adopters or not, e-commerce actors, traditional users - can benefit from and access a democratic and permissionless environment, where economic and governance fundamentals are used in order to drive the best possible optimization and thus the proper execution of several types of exchanges, a financial gain and decision power on the part of the actors involved and to optimize transactional behaviors within the system. In a decentralized model, the best way to ensure a smooth operation and a stable environment for its users without acting as a third party during the various interactions is to set up the necessary mechanisms to incentivize the actors, thus raising their respective personal interests and algorithmically penalizing them in a significant way in the case of dishonest behavior. This is made possible by the use of the distributed ledger offered by blockchain technologies, first via Polygon Chain and then the Lyzis Chain with an \textit{EVM} configuration (see Section \ref{mvp} for more details). As a result, all the actors concerned are then able to contribute at their own scale to the development and efficient functioning of the platform.

Lyzis Marketplace operates efficiently and securely as an untrusted ecosystem, where arbitrarily users are allowed to join a global network to sell/shop online; they are incentivized to participate in the construction and design of Lyzis in a beneficial way. Actors can commonly interact in an environment where trust is built-in even for first time buyers, people unfamiliar with blockchains, and people disappointed by traditional marketplaces. In our view, facilitating the key functions of marketplaces, decentralization and the use of blockchain technology will, if successful, complement and compete with today's dominant centralized protocols.

\subsection{Comprehensive Description}

More concretely, a seller can list online on Lyzis Marketplace his physical assets for sale whether they are old, used or new. An item can only be sold on the platform if it can be sent in package format to a buying user (e.g., no categories for cars, large objects). 
This limits the cases of possible physical interactions, hypothetically increasing the need for trust and the perception of potential and additional risks.
The entire operation is carried out in P2P and totally decentralized without having to go through a TTP to get access to the exchange platform nor to question the seriousness of the peer user. 
This allows to avoid the classical problems associated to centralized structures and, thanks to the use of a smart contract as a classic escrow contract, to take advantage of all the benefits inherent to the blockchain.
Thanks to the blockchain technology, which is the core component underlying the Lyzis Marketplace:

\begin{itemize}
    \item Any buyer can directly contact any seller in a decentralized, transparent and secure way where both will trust each other without the need for a TTP (i.e., none of them can and has a personal interest in trying to harm the other for several reasons, the main one being the securing and blocking of funds in the smart contract and the non-possibility for each of the actors to access them - each actor relies on the system to ensure integrity).

    \item Any buyer can confirm the identity of any seller without the need for a TTP - \textit{seller authentication} (i.e., the buyer is sure that the wallet of the seller from whom he wants to buy an asset is active and operating on the blockchain, thus being able to access, among other things and in a transparent way if necessary, its liabilities within the Lyzis' network).
    
    \vspace{+0.05cm}
    
    \item Similarly, any seller can confirm the identity of any buyer without the need for a TTP - \textit{buyer authentication} (i.e., the seller is sure that the wallet of the buyer to whom he has to sell and send an asset is active, authenticated within the network and runs on the blockchain, thus being able to access, among other things and in a transparent way if necessary, its liabilities within the Lyzis' network).
    
    \vspace{+0.05cm}
    
    \item A buyer can pay an item to a seller only in digital assets and more specifically, depending on the seller's choice, in LZS or LZDC (see Section \ref{assetslyzis}). The seller can then receive payment for his sold assets in the desired digital asset (presumably stable based on his rationality) and a dynamic conversion is set up to automatically convert LZS \textbf{$\leftrightarrow$} LZDC according to the predefined choice to facilitate exchanges.
    
    \vspace{+0.05cm}
    
    \item Any user can then claim to be rewarded in LZSP (classic ERC20 governance token) for each beneficial participatory action leading to the efficient and functioning of the marketplace, e.g., validation of the receipt and/or shipment of an item where applicable, post-exchange review of the other party (i.e., scoring, comments) (see Fig.\ref{Fig3} and/or \cite{Lyzistokenomics}). This governance token is then used to incentivize agents within the ecosystem to a positive and honest behavior while making them gain decision-making and economic power (i.e., a dynamic conversion is also set up to automatically convert LZS \textbf{$\leftrightarrow$} LZDC) and decision-making power in order to allow each one to participate at its own scale, among others, to the Lyzis decentralized governance.
\end{itemize}

In summary, each category of actors involved in the Lyzis Marketplace takes advantage of all the specific benefits of using blockchain technology in their various interactions. 
Also, the autonomy and independence of Lyzis are reinforced by several features that can reward its members according to their participation level (see Section \ref{techfeat&spe}).
Fig.\ref{Fig1} shows a simple graphical representation of the functioning of an exchange - with a branded watch as physical asset - on the Lyzis Marketplace in which two parties, buyer and seller, are involved and connected in a decentralized way (i.e., without the intervention of a TTP) with a delivery assured by traditional postal distribution channels.

Thereafter, Section \ref{functionaldescription} provides more details on how the algorithm and the smart contract dedicated to handle physical asset exchanges work. 
Section \ref{conflict} presents more details about the management of decentralized conflicts that may arise within the Lyzis Marketplace.

\vspace{+0.2cm}

\begin{figure}[h!]
\includegraphics[width=0.9\textwidth]{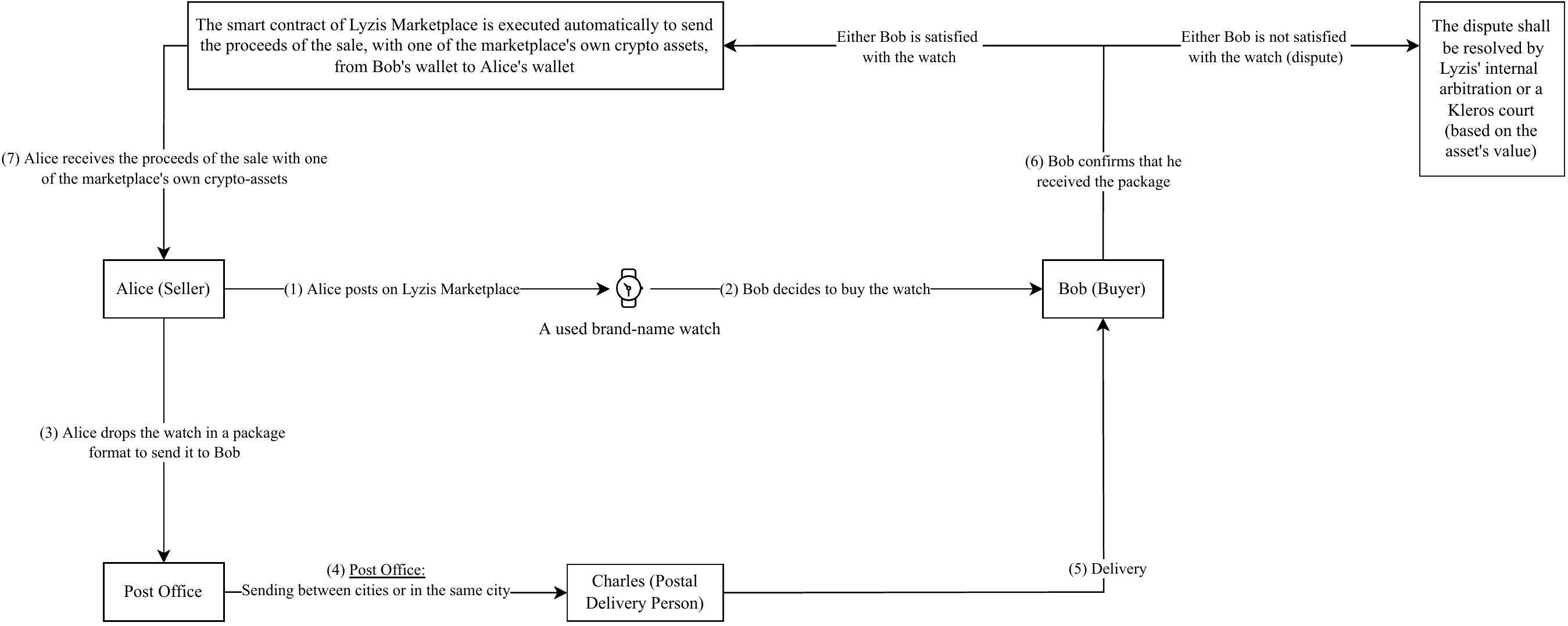}
\center
\caption{How a P2P exchange of physical assets on Lyzis Marketplace works using a traditional postal service.}
\label{Fig1}
\end{figure}

\subsection{Objectives}

Lyzis is proposing an integrated solution to solve the problems facing  traditional, centralized commerce marketplaces. Trust and security are built in by design, by  using decentralized consensus blockchain technology. Giving participants the power to engage in an open marketplace without a centralized entity gives them a sense of economic freedom and empowerment.

Aside from blockchain technology's cryptographic means of data encryption, censorship resistance, fault-tolerance, immutability, security and transparency, blockchain technology has enabled decentralized transaction of goods and services with a digitally generated currency that allows everyone to participate without discrimination. Decentralization and censorship resistance as well as data integrity, untrustedness and optimal security make blockchain technology an ideal tool to achieve the target of creating a decentralized marketplace for the physical asset exchanges.
Consequently, the use of a distributed ledger technology as an underlying infrastructure eliminates the traditional finance model of transactions where intermediaries determine the trust of data and transactions. 
Lyzis is building a decentralized environment where everyone can participate, interact and transact; hence revolutionizing the way today's commercial transactions are executed.  

\subsubsection{Use-Cases}

Blockchain technology's benefits are not enough on their own to justify the intrinsic value and usefulness of a decentralized venture. 
To be concrete and beyond the simple use of blockchain and its associated advantages, we shall describe the main use cases of Lyzis Marketplace as: solving trust and adoption issues.

\vspace{-0.1cm}

\paragraph{Solving Trust Issue.}

As previously stated, the main and central point of Lyzis is to overcome the trust issues experienced on centralized commercial platforms of all types by leveraging mainly the blockchain technology and applying several game theoretical based mechanisms. This builds trust that can be attributed not to the individual participants, but to the system that enables the operation and support of any predefined activity in a reliable and transparent way.
In this instance, blockchain technology, consolidated by smart contracts, allows to bring autonomy, and guarantee the security of the on-chain data and integrity of the business logic execution defined in smart contracts. 
Within a Dapp and more especially within a decentralized marketplace, there is a critical need for the implementation of an additional system allowing to reduce the perception of risk that can be observed by different users, thereby increasing the level of perceived trust towards the network. 
Decentralization leads to a well-defined paradigm here: the trustless needs trust - in other words, the proper functioning of the system and P2P operations can only be ensured by the provision of solid trust by the different participants towards the network so that they no longer have to trust each other as they are completely unknown.
\\
We therefore build an integrated trust system (see Section \ref{techfeat&spe}) with the contribution of several components to hinder deceptive actors while encouraging honest ones.
The trust system  operates as an underlying decentralized service providing a feedback mechanism for end users and maintaining trust relationships between them in the Lyzis ecosystem.

\paragraph{Solving Adoption Issue.}

We strongly believe that besides solving the trust problems usually faced by mainstream users on centralized marketplaces, it is possible for Lyzis Marketplace to help solve an adoption/access issue to the blockchain field by allowing an initial interaction with the space and the ability for an end-user to keep exploring it afterwards.
This aspect should be considered and highlighted by all decentralized projects during their design to allow an easier and faster mass adoption within the blockchain space to support its development and openness.
\\
For Lyzis Marketplace, let us take the case of a user (Bob) who has already sold goods of any kind on traditional hierarchical platforms but has no experience in the field of blockchain. Bob has heard about the blockchain ecosystem and/or digital assets, but is confused about where to start and does not want to spend personal fiat as he finds this area far too risky, mainly because of their reputation of volatility\footnote{Knowing the well-known potential volatility of crypto-currencies, the risk associated with fluctuations remains a substantial barrier to entry into the blockchain space for a conventional user. This translates into the expression/increase of anxiety that can be expressed by some users towards prospective online purchases.}.
He then has the opportunity to list an item he no longer has any utility for on Lyzis Marketplace, to sell it, and thus to be remunerated in digital assets (stable or volatile depending on his choice) while taking advantage of the security and other benefits brought by the use of the blockchain in order to be able to discover and begin to apprehend decentralization.

Lyzis Marketplace therefore provides cost-effective access to the web3. 
Additional uses are set out in Section \ref{future} on potential further implementations within the platform.

\subsection{Architectural Design and Functional Description}\label{functionaldescription}

This section provides a technical description of the architecture of Lyzis Marketplace.

Fig.\ref{Fig2} shows the sequence diagram of an asset exchange within the proposed platform among different actors and components, i.e., the smart contract of Lyzis Marketplace, Alice (seller), the website of Lyzis Marketplace, Bob (buyer), and Charles (delivery man). 

These actors and components are bound here by a trust relationship. A trust relationship is a relationship involving several entities (in this case, people and/or software components). The entities in a relationship trust each other to have or not to have certain properties\footnote{In the case of Lyzis Marketplace, the properties are represented by the set of mechanisms implemented within the trust system to essentially prevent dishonest behaviors and inefficiencies (see Section \ref{techfeat&spe}).} (the "trust assumptions"). If the trusted entities satisfy these properties, then they are trustworthy \cite{trustrelationship}.
Considering also the notion of online trust defined in Section \ref{over}, where in the case of a P2P system a user trusts the system more than the adversary participants, we can define in Tab.\ref{Tab1} the actors and components involved in a physical asset exchange on Lyzis Marketplace and their respective roles. We can also formalize the trust assumptions (1) and (2).

\begin{table}[h!]
\resizebox{\textwidth}{!}{%
\begin{tabular}{|l|l|l|}
\hline
\rowcolor[HTML]{EFEFEF} 
\multicolumn{1}{|c|}{\cellcolor[HTML]{EFEFEF}\textbf{Actor and/or component involved}} & \multicolumn{1}{c|}{\cellcolor[HTML]{EFEFEF}\textbf{Definition and/or role}} & \multicolumn{1}{c|}{\cellcolor[HTML]{EFEFEF}\textbf{Trust assumptions}} \\ \hline
\rowcolor[HTML]{FFFFFF} 
Smart contract of Lyzis Marketplace & \begin{tabular}[c]{@{}l@{}}The Lyzis marketplace's smart contract is an \\ encoded, self-executing code whose terms of\\ agreement between buyer and seller are auto-\\ matically verified and executed via the com-\\ puter network. Its defined role is to simplify\\ business and trade of physical assets between\\ unknown people (Alice and Bob) without ha-\\ ving to trust each other and rely on the TTP\end{tabular} &  \\
\rowcolor[HTML]{EFEFEF} 
{\color[HTML]{000000} Alice (seller)} & {\color[HTML]{000000} \begin{tabular}[c]{@{}l@{}}Alice is a seller on Lyzis Marketplace, selling\\ a physical asset (defined as "item I" in Fig.\ref{Fig2}\\ and \ref{Fig3})\end{tabular}} & {\color[HTML]{000000} \begin{tabular}[c]{@{}l@{}}(1) We assume that Alice trusts (i) the smart contract\\ of Lyzis Marketplace because its operation is\\TTP-free and autonomous, (ii) the Lyzis Marketplace\\website because the operation of the exchange relies\\on the use of the smart contract, and (iii) Bob (buyer)\\because he satisfies the predefined properties, e.g., his\\account is active and validated and/or he maintains a\\good reputation level within the platform\end{tabular}} \\
{\color[HTML]{000000} Lyzis Marketplace website} & {\color[HTML]{000000} \begin{tabular}[c]{@{}l@{}}The Lyzis Marketplace website is an online\\interface allowing users to interact with the\\Lyzis Marketplace smart contract and to \\consequently perform exchanges\end{tabular}} & {\color[HTML]{000000} } \\
\rowcolor[HTML]{EFEFEF} 
{\color[HTML]{000000} Bob (buyer)} & {\color[HTML]{000000} \begin{tabular}[c]{@{}l@{}}Bob is a seller on Lyzis Marketplace, buying\\a physical asset (defined as "item I" in Fig.\ref{Fig2}\\and \ref{Fig3})\end{tabular}} & {\color[HTML]{000000} \begin{tabular}[c]{@{}l@{}}(2) We assume that Bob trusts (i) the smart contract\\of Lyzis Marketplace because its operation is\\TTP-free and autonomous, (ii) the Lyzis Marketplace\\ website because the operation of the exchange relies\\on the use of the smart contract, and (iii) Alice (seller)\\because she satisfies the predefined properties, e.g.,\\her account is active and validated by the required\\stacking and/or she maintains a good reputation level\\ within the platform\end{tabular}} \\
{\color[HTML]{000000} Charles (delivery man)} & {\color[HTML]{000000} \begin{tabular}[c]{@{}l@{}}Charles is a delivery person employed by a\\pre-defined external postal service. His defi-\\ned role is to deliver the package according\\to the conditions established by the relevant\\post office\end{tabular}} & {\color[HTML]{000000} } \\ \hline
\end{tabular}%
}
\caption{Definition and role of the actors and/or components involved in an exchange on the Lyzis Marketplace, as well as the key trust assumptions.}
\label{Tab1}
\end{table}

\begin{figure}
\centering
\includegraphics[width=0.845\textwidth]{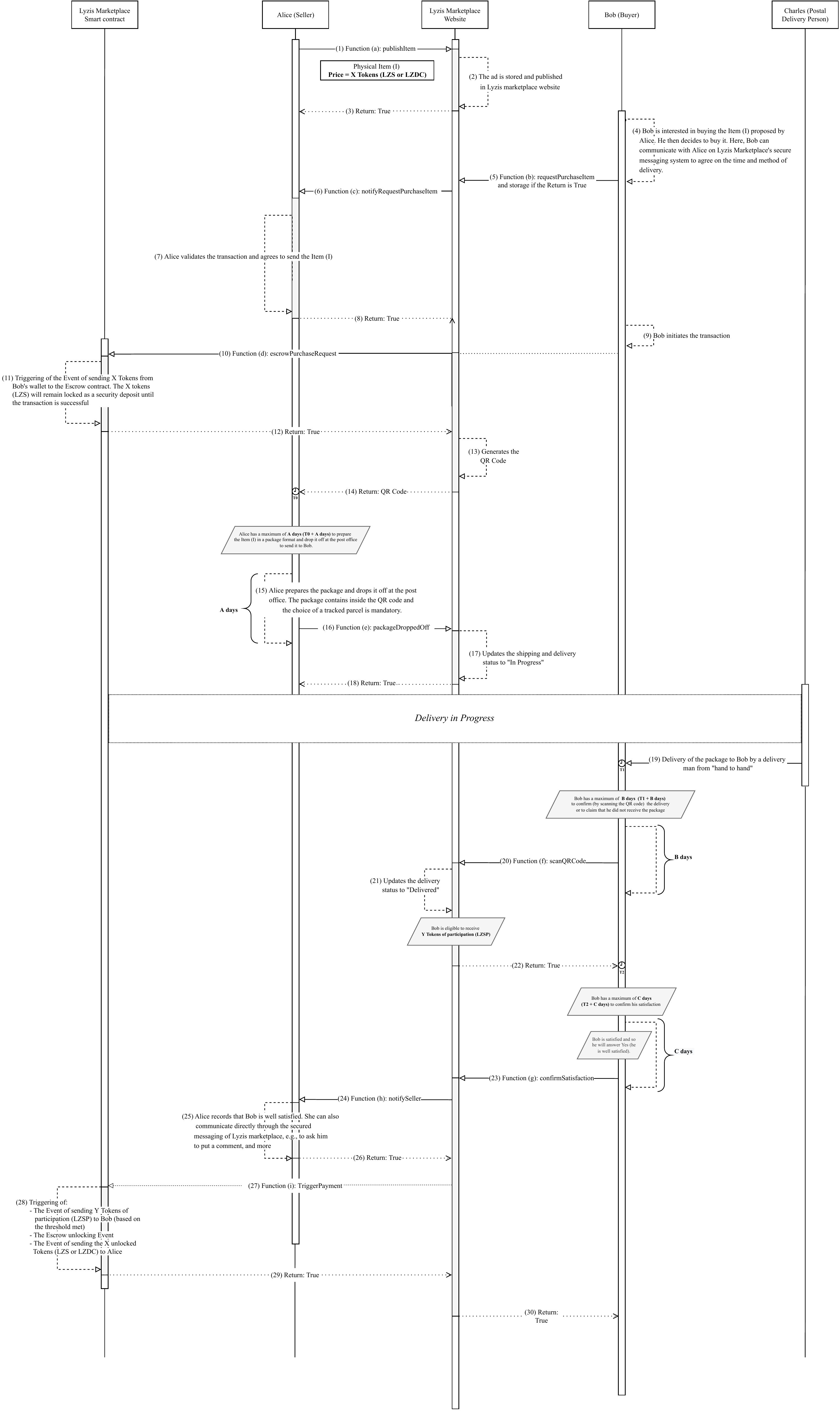}
\caption{UML diagram of a sample workflow of the proposed marketplace (exchange related).}
\label{Fig2}
\end{figure}

\subsubsection{\textit{Function Details (i)}}
The following Tab.\ref{Tab2} sums up and describes the various elements and functions shown in Fig.\ref{Fig2}.

\begin{table}[!h]
    \centering
    \begin{threeparttable}
    \captionof{table}{Functions description of the sequence diagram (Fig.\ref{Fig2}). \label{Tab2}}
    \begin{tabularx}{\textwidth}{XXX}
    Element(s)/Function(s) \tnote{*}  & Specification(s) & Reference\\
     \midrule\midrule

    publishItem    
    & The seller (here Alice) places an ad on the marketplace for an item (I) that she wishes to sell.
    &  \makecell{[1](a)} \\
    \cmidrule(l  r ){1-3}
    
    The ad is stored and published in Lyzis Marketplace website
    &   The ad in question of item (I) is placed on Lyzis Marketplace, registered (provided that the seller's wallet is valid) and visible to all participants.
    & \makecell{[2]} \\ 
    \cmidrule(l r ){1-3}
    
    Return: True 
    & Action of informing the seller (Alice) of the good deposit of his item (I) on the marketplace.
    & \makecell{[3]} \\ 
    \cmidrule(l r ){1-3}
    
    Bob is interested in buying the Item (I) proposed by Alice. He then decides to buy it. Here, Bob can communicate with Alice on Lyzis Marketplace's secure messaging system to agree on the time and method of delivery
    & The buyer (here Bob) sees and wants to buy the item (I). He communicates with the seller (here Alice) to agree on the final terms of the exchange. 
    & \makecell{[4]} \\ 
    \cmidrule(l r ){1-3}
    
    requestPurchaseItem
    & The buyer (Bob) indicates to the seller (Alice) that he wants to buy item (I) by formalizing a shipping request on the marketplace. If true, the previous shipping request / purchase request is stored on the marketplace's website.
    & \makecell{[5](b)} \\ 
    \cmidrule(l r ){1-3}
    
    notifyRequestPurchaseItem 
    & The seller (Alice) is informed by Lyzis Marketplace of the potential sale and she agreed to the final terms of the exchange with Bob.
    & \makecell{[6](c)} \\
    \cmidrule(l r ){1-3}
    
        Alice validates the transaction and agrees to send the Item (I)
    & The seller (Alice) validates the transaction. The conditions are definitely set (e.g., date, shipping method).
    & \makecell{[7]} \\ 
    \cmidrule(l r ){1-3}

    \end{tabularx}
    \end{threeparttable}
\end{table}


\begin{table}[!h]
    \centering
    \begin{threeparttable}
    \begin{tabularx}{\textwidth}{XXX}
    
    Return: True 
    & Information action of the Lyzis Marketplace concerning the good agreement of the parties (seller and buyer) on the terms of the exchange.
    & \makecell{[8]} \\ 
    \cmidrule(l r ){1-3}
    
    Bob initiates the transaction
    & Bob initiates the transaction by signing with his private key via the proposed interface of the Lyzis Marketplace.
    & \makecell{[9]} \\ 
    \cmidrule(l r ){1-3}
    
    escrowPurchaseRequest 
    & Interaction from the marketplace to the platform's smart contract: the website instantiates the smart contract and asks it to start the sales process for item (I).
    & \makecell{[10](d)} \\ 
    \cmidrule(l r ){1-3}
    
    Triggering of the Event of sending  X Tokens from Bob's wallet to the Escrow contract. The X Tokens (LZS) will remain locked as a security deposit until the transaction is successful
    & Once the buyer (Bob) has signed the transaction and deposited in the smart contract the value of the item (I) he wants to buy, these become inaccessible by any of the parties (buyer/seller/platform) and are locked until the good realization of the exchange (the sending to the seller (Alice)) in case this one is realized correctly.
    & \makecell{[11]} \\ 
    \cmidrule(l r ){1-3}
    
    Return: True 
    & Information action from the smart contract to the Lyzis Marketplace regarding the successful deposit of funds from the buyer (Bob) for the exchange.
    & \makecell{[12]} \\ 
    \cmidrule(l r ){1-3}
    
    Generates the QR Code
    & Alice may be able at this stage to request the generation of a QR code on the marketplace that will be integrated inside the package and set to be scanned by the receiving party (Bob), among other things to confirm the good reception (preventive measure). If True, the QR code requested in function by the seller (Alice) is then initiated from his interface.
    & \makecell{[13]} \\ 
    \cmidrule(l r ){1-3}
    
    \end{tabularx}
    \caption{\textit{Continued from previous page}.}
    \end{threeparttable}
\end{table}

 
\begin{table}[!h]
    \centering
    \begin{threeparttable}
    \begin{tabularx}{\textwidth}{XXX}
    
    Return: QR Code 
    & The QR code initiated in the previous function and then accessible from the seller's interface (Alice) to be deposited inside the package ready for shipment.
    & \makecell{[14]} \\ 
    \cmidrule(l r ){1-3}
    
    Alice prepares the package and drops it off at the post office. The package contains inside the QR code and the choice of a tracked parcel is mandatory 
    & The seller (Alice) has then (T0 + A days) to prepare the package of the item (I) with the QR code inside, to deposit it at a post office with the mandatory choice of a tracked package, and to validate its sending to the buyer (Bob) on the marketplace from its interface within the given time.
    & \makecell{[15]} \\ 
    \cmidrule(l r ){1-3}
    
    packageDroppedOff 
    & The package is well prepared and deposited by the seller (Alice) within the time limit as agreed in the previous function.
    & \makecell{[16](e)} \\ 
    \cmidrule(l r ){1-3}
    
    Updates the shipping and delivery status to "In Progress" 
    & The seller (Alice) then confirms the effective shipment of the package containing item (I) and also submits its assigned tracking number - from its dedicated interface on Lyzis Marketplace.
    & \makecell{[17]} \\ 
    \cmidrule(l r ){1-3}
    
        Return: True 
    & The shipment by the seller (Alice) is well registered and validated by the marketplace, the buyer (Bob) is then informed of the shipment and has the tracking of the package containing the item (I).
    & \makecell{[18]} \\ 
    \cmidrule(l r ){1-3}
    
    \end{tabularx}
    \caption{\textit{Continued from previous page--}.}
    \end{threeparttable}
\end{table}


\begin{table}[!h]
    \centering
    \begin{threeparttable}
    \begin{tabularx}{\textwidth}{XXX}

    Delivery of the package to Bob by a delivery man "from hand to hand" 
    &  The parcel then arrives effectively at the buyer's home (Bob) and is dropped off by the delivery person linked to the postal service chosen for the shipment by the seller (Alice). 
    & \makecell{[19]} \\ 
    \cmidrule(l r ){1-3}
    
    scanQRCode 
    & The buyer (Bob) has a time limit of (T1 + B days) to open the package received, scan the QR code available inside it and thus definitively confirm its receipt.
    & \makecell{[20](f)} \\ 
    \cmidrule(l r ){1-3}
    
    Updates the delivery status to "Delivered" 
    & Update event on the marketplace regarding the shipping status of the package being previously "in progress" to "delivered". The seller (Alice) is then aware of the good reception of the item (I) by the buyer (Bob).
    & \makecell{[21]} \\ 
    \cmidrule(l r ){1-3}
    
    Return: True 
    & Information action from the Lyzis Marketplace about the correct update of the status of the package ("in progress" to "delivered"). The buyer (Bob) then knows that the seller (Alice) is aware of its good reception.
    & \makecell{[22]} \\ 
    \cmidrule(l r ){1-3}
    
    confirmSatisfaction 
    & The buyer (Bob) has a time limit of (T2 + C days) to confirm his satisfaction (e.g., status of item (I) as agreed). Assuming he is satisfied, then he answers "Yes" and his answer is transmitted and recorded by the Lyzis Marketplace.
    & \makecell{[23](g)} \\ 
    \cmidrule(l r ){1-3}

    \end{tabularx}
    \caption{\textit{Continued from previous page}.}
    \end{threeparttable}
\end{table}


\begin{table}[!h]
    \centering
    \begin{threeparttable}
    \begin{tabularx}{\textwidth}{XXX}
    
    Function (i): notifySeller 
    & Action of informing the seller (Alice) by the Lyzis Marketplace about the satisfaction level obtained by the buyer (Bob) following the exchange.
    & \makecell{[24](h)} \\ 
    \cmidrule(l r ){1-3}

    Alice records that Bob is well satisfied. She can also communicate directly through the secured messaging of Lyzis Marketplace, e.g., to ask him to put a public comment
    & The seller (Alice) is informed of the good satisfaction of his interlocutor by the Lyzis Marketplace. They can also discuss it through the messaging system.
    & \makecell{[25]} \\ 
    \cmidrule(l r ){1-3}
    
    Return: True 
    & Information action of the Lyzis Marketplace concerning the good reception from the seller (Alice) of the satisfaction level of the buyer (Bob).
    & \makecell{[26]} \\ 
    \cmidrule(l r ){1-3}
    
        TriggerPayment 
    & The Lyzis Marketplace registers this information and interacts with the smart contract, having the necessary conditions to release the funds initially inserted in it by the buyer (Bob) and to guarantee the payment of the item (I) to the seller (Alice).
    & \makecell{[27](i)} \\ 
    \cmidrule(l r ){1-3}
    
    Triggering of: The Event of sending Y Tokens of participation (LZSP) to Bob (based on the threshold met), The Escrow unlocking Event, The Event of sending the X unlocked Tokens (LZS or LZDC) to Alice 
    & The smart contract, provided that the necessary conditions are met, then \textit{(i)} unlocks the funds of the buyer (Bob) blocked and initially planned for the payment of the seller (Alice) being in LZS or LZDC, \textit{(ii)} unlocks the LZSP participation tokens intended to reward all the honest and contributing behaviors to the good functioning of the exchange carried out, and thus sends these tokens to the various parties concerned.
    & \makecell{[28]} \\ 
    \cmidrule(l r ){1-3}

    \end{tabularx}
    \caption{\textit{Continued from previous page}.}
    \end{threeparttable}
\end{table}

\begin{table}[!h]
    \centering
    \begin{threeparttable}
    \begin{tabularx}{\textwidth}{XXX}
    
    Return: True 
    & Information action from the smart contract to the Lyzis Marketplace regarding the successful release of funds and the various allocations and transfers mentioned in the previous section.
    & \makecell{[29]} \\ 
    \cmidrule(l r ){1-3}
    
    Return: True 
    & Information action of the seller by the Lyzis Marketplace and to the buyer (Bob) about the successful operation.
    & \makecell{[30]} \\ 
    \midrule\midrule    
    \end{tabularx}
    \begin{tablenotes}
\item[*] Throughout the tables, \textit{References} defined as \textit{[x](y)} are then functions identified as such in Fig.\ref{Fig2}, while any other elements defined as \textit{[x]} refer to any interactions between the different "\textit{actors}" involved (smart contract of Lyzis Marketplace, Alice (Seller), Lyzis Marketplace Website, Bob (Buyer) and Charles (Deliverer employed by the Post Office)).
\end{tablenotes}
\caption{\textit{Continued from previous page}.}
\end{threeparttable}
  \end{table}

  \clearpage

All time-related data mentioned in the previous tables will then be defined at a later stage.

 Fig.\ref{Fig3} shows a precise description of the functioning of the algorithm integrated within the Lyzis Marketplace for an asset exchange. In Fig.\ref{Fig3}, we assume that the cases of "\textit{right(s) of return}" and "\textit{return for non-conformity}" according to the applicable regulations and consumer protection laws are covered by the cancellation functions\footnote{We also assume that the seller is made aware of the applicable return rights periods (cancellation periods) before, during, and after the successful execution of the exchange.} C3, C4, and C5 (see Tab.\ref{Tab11}, \ref{Tab12} and \ref{Tab13}).
However, the time variables (i.e., T0, T1, T1', T2, T3, and A, B, B', C, D, E, and F days) being undefined, their respective attributions shall be studied later based on the country or countries concerned.
  
\begin{figure}[hp]
\centering
\includegraphics[width=0.65\textwidth]{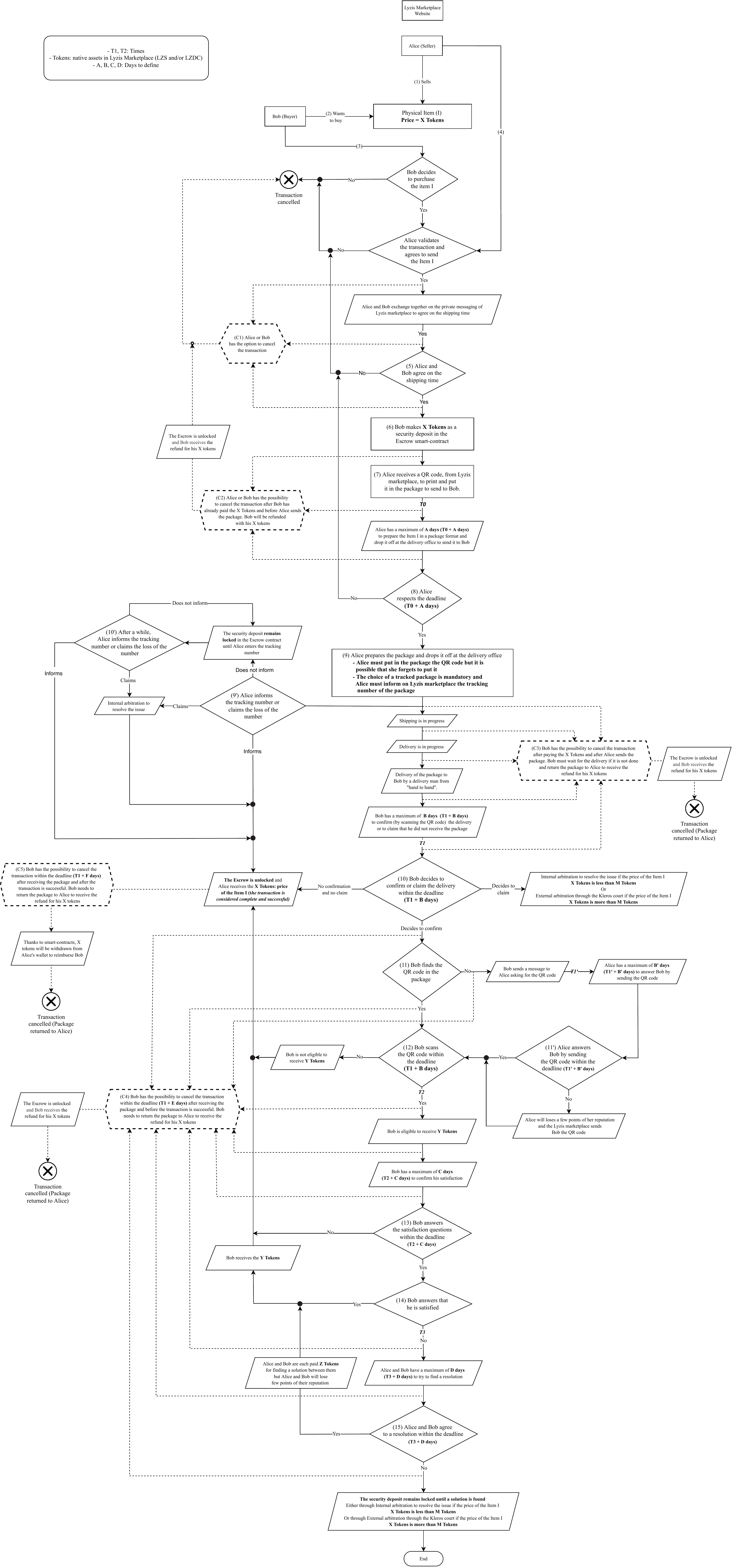}
\caption{Technical description of the Lyzis Marketplace algorithm representing step by step the realization of a P2P object exchange.}
\label{Fig3}
\end{figure}
\begin{center}
\end{center}

\newpage

\subsubsection{\textit{Function Details (ii)}}
The following Tab.\ref{Tab8} sums up and describes the various elements/functions shown in Fig.\ref{Fig3}.
\begin{table}[!hp]\label{Tab8}
    \centering
    \begin{threeparttable}
    \captionof{table}{Functions description of the operating algorithm (Fig.\ref{Fig3}). \label{Tab8}}
    \begin{tabularx}{\textwidth}{XXX}
    Function(s)  & Specification(s) & Reference\\
     \midrule\midrule
    
    \textbf{\textit{Action functions}}
    &
    & \textbf{(1) to (15)} \\
    \cmidrule(l  r ){1-3}

    Sells
    & Alice sells a physical item (I) for X tokens.
    & \makecell{[1]} \\
    \cmidrule(l  r ){1-3}
    
    Bob Wants to Buy and decides to purchase the item (I)
    & Bob notices item (I) on the platform and decides to buy it.
    & \makecell{[2,3]} \\ 
    \cmidrule(l r ){1-3}
    
    Alice validates the transaction and agrees to send the item (I)
    & Alice is informed of the purchase request, and validates the transaction, committing at this stage to preparing the package. If this is not done, the transaction is cancelled.
    & \makecell{[4]} \\ 
    \cmidrule(l r ){1-3}
    
    Alice and Bob agree on the shipping time
    & After exchanging via the platform's messaging system, Alice and Bob agree on a sending time. If this is not the case, the transaction is cancelled.
    & \makecell{[5]} \\ 
    \cmidrule(l r ){1-3}
    
    Bob makes X Tokens as a security deposit in the Escrow smart-contract
    & Bob deposits the value of item (I), the X tokens, in the exchange's smart contract, considered as a security deposit. If this is not done, the transaction is cancelled.
    & \makecell{[6]} \\ 
    \cmidrule(l r ){1-3}
    
    Alice receives a QR code, from Lyzis Marketplace, to print and put it in the package to send to Bob and she respect the deadline (T0 + A days)
    & Alice retrieves the QR code generated by the platform to be printed and integrated inside the package. She has a maximum of A days (T0 + A days) to put the item (I) in package format with the QR code, deposit it at the post office according to the instructions received. If this is not the case, the transaction is cancelled.
    & \makecell{[7,8]} \\
    \cmidrule(l r ){1-3}

    \end{tabularx}
    \end{threeparttable}
\end{table}

\begin{table}[!h]
    \centering
    \begin{threeparttable}
    \begin{tabularx}{\textwidth}{XXX}
    
    Alice prepares the package and drops it off at the delivery office. Alice must put in the package the QR code but it is possible that she forgets to put it. 
    The choice of a tracked package is mandatory and Alice must inform on Lyzis Marketplace the tracking number
    & Alice prepared and dropped off the package as instructed at the post office. She can forget to insert the QR code inside, which implies that \textit{(i)} Bob will not be able to scan it and get paid in LZSP (see Section \ref{ecoincentive}) so he gets them by default and Alice won't be awarded any more, and \textit{(ii)} she will have one less evidence in case she needs to appeal to an external court (see Section \ref{conflict}) for any dispute (i.e., it's not in her interest to forget).
    & \makecell{[9]} \\ 
    \cmidrule(l r ){1-3}
    
    Alice informs the tracking number or claim the loss of the number
    & Alice informs the tracking number or informs of its loss. The implications are the same as for [9]. If she informs and the delivery is successful and validated by Bob, the escrow is released and Alice receives the X Tokens. If this is not the case, i.e., she does not inform the tracking number and additionally the transaction is not validated by Bob (receipt), the security deposit remains locked until this is the case, and it may be possible for Bob to go to arbitration.
    & \makecell{[9']} \\ 
    \cmidrule(l r ){1-3}
    
    After a while, Alice informs the tracking number or claims the loss of the number
    & If Alice, after not informing the tracking number, declares her loss, the case shall be resolved essentially by internal arbitration. If she keeps on not informing the tracking number without any news, and additionally the transaction is not validated by Bob (recept), the security deposit remains locked until this is the case, and it may be possible for Bob to go to arbitration.
    & \makecell{[10']} \\ 
    \cmidrule(l r ){1-3}
    
    \end{tabularx}
    \caption{\textit{Continued from previous page}.}
    \end{threeparttable}
\end{table}

\begin{table}[!h]
    \centering
    \begin{threeparttable}
    \begin{tabularx}{\textwidth}{XXX}
    
    Bob decides to confirm or claim the delivery within the deadline (T1 + B days)
    & After the shipment and delivery have, or have not, taken place, Bob has a maximum of B days (T1 + B days) to confirm the reception of the package, mainly by scanning the QR code or by validating through the interface. He can also, if not, decide to claim. If he does not confirm and does not claim, the security deposit is released, Alice gets the X tokens, and the transaction is successful by default. If he claims, the issue shall be resolved by the designated arbitration court (see Section \ref{conflict}).
    & \makecell{[10]} \\ 
    \cmidrule(l r ){1-3}
    
    Bob finds the QR code in the package
    & Bob decides to confirm. He must find the QR code in the package. If he can't find it, he can initially send a message to Alice to ask her for it (i.e., in case she has forgotten). Alice has B' days (T1' + B' days)\footnote{Here, we note that: T1 + B days $>$ T1' + B' days.} to answer Bob.
    & \makecell{[11]} \\ 
    \cmidrule(l r ){1-3}
    
    Alice answers Bob by sending the QR code within the deadline
    & Alice must answer Bob and send him the QR code. If she doesn't answer within the deadline (T1' + B' days), Alice loses reputation points and the marketplace sends the QR code to Bob.
    & \makecell{[11']} \\ 
    \cmidrule(l r ){1-3}
    
    Bob scans the QR code within the deadline (T1 + B days)
    & Bob gets the QR code. If he scans it in the deadline, T1 + B days, he is eligible to receive LZSP. If not, he is not and the transaction is set as successful by default, the security deposit is released. He also has T2 + C days to confirm his satisfaction (see \cite{Lyzistokenomics} for more details on LZSP distribution mechanisms).
    & \makecell{[12]} \\ 
    \cmidrule(l r ){1-3}
    
    \end{tabularx}
    \caption{\textit{Continued from previous page}.}
    \end{threeparttable}
\end{table}

\begin{table}[!h]
    \centering
    \begin{threeparttable}
    \begin{tabularx}{\textwidth}{XXX}
    
    Bob answers the satisfaction questions within the deadline (T2 + C days)
    & If Bob does not answer the satisfaction questions within the T2 + C days deadline, Bob is not eligible for LZSP (see \cite{Lyzistokenomics} for more details on LZSP distribution mechanisms) and the security deposit is released, the transaction is considered successful.
    & \makecell{[13]} \\ 
    \cmidrule(l r ){1-3}
    
    Bob answers that he is satisfied
    & If Bob answers and declares that he is not satisfied, Bob and Alice must find a solution in a maximum of D days (T3 + D days). If Bob answers and declares to be satisfied, he receives Y tokens (see \cite{Lyzistokenomics} for more details on LZSP distribution mechanisms) and the security deposit is released, the transaction is considered successful.
    & \makecell{[14]} \\ 
    \cmidrule(l r ){1-3}
    
    Alice and Bob agree to a resolution within the deadline
    & If Bob and Alice have achieved a resolution within the deadline (T3 + D days), Alice and Bob receive Z tokens (see \cite{Lyzistokenomics} for more details on LZSP distribution mechanisms), Alice loses some reputation points and the security deposit is released, the transaction is considered successful. If they fail to find a solution, the security deposit remains locked until a solution is found according to the relevant arbitration (see Section \ref{conflict}).
    & \makecell{[15]} \\ 
    
     \midrule\midrule
    
    \textbf{\textit{Cancellation functions}}
    & 
    & \textbf{(C1) to (C5)} \\ 
    \cmidrule(l r ){1-3}
    
    Alice or Bob has the option to cancel the transaction
    & Possibility for both parties (i.e., buyer and seller) to cancel the transaction (from [4] to [5] and from [5] to [6]). In this instance, the transaction is cancelled.
    & \makecell{[C1]} \\ 
    \cmidrule(l r ){1-3}
    
    \end{tabularx}
    \caption{\textit{Continued from previous page}.}
    \label{Tab11}
    \end{threeparttable}
\end{table}

\begin{table}[!h]
    \centering
    \begin{threeparttable}
    \begin{tabularx}{\textwidth}{XXX}

    Alice or Bob has the possibility to cancel the transaction after Bob has already paid the X tokens and before Alice sends the package. Bob will be refunded with his X tokens
    & Possibility for both parties (i.e., buyer and seller) to cancel the transaction (from [6] to [7] and from [7] to [8]) once Bob has paid and before Alice has sent the package. In this instance, the transaction is cancelled, the escrow is released and Bob is refunded with his X tokens.
    & \makecell{[C2]} \\
    \cmidrule(l r ){1-3}

    Bob has the possibility to cancel the transaction after paying the X Tokens and after Alice sends the package. Bob must wait for the delivery if it is not done and return the package to Alice to receive the refund for his X Tokens
    & Possibility for Bob (i.e., the buyer) to cancel the transaction (from [9] to [10]) once he has paid and after Alice has sent the package (\textit{right of return} applicable). In this instance, the transaction is cancelled, the package returned to Alice, the escrow is released and Bob is refunded with his X tokens.
    & \makecell{[C3]} \\ 
    \cmidrule(l r ){1-3}
    
    Bob has the possibility to cancel the transaction within the deadline (T1 + E days) after receiving the package and before the transaction is successful. Bob needs to return the package to Alice to receive the refund for his X tokens
    & Possibility for Bob, within the T1 + E days deadline, to cancel the transaction (from [10] to [11], [11] to [11'], [11] to [12], [12] to [13], and [14] to [15]) once he has received the package and before the transaction is considered complete and successful (\textit{right of return} and \textit{return for non-conformity} applicable). In this instance, the transaction is cancelled, the package returned to Alice, the escrow is released and Bob is refunded with his X tokens.
    & \makecell{[C4]} \\ 
    \cmidrule(l r ){1-3}
    
    \end{tabularx}
    \caption{\textit{Continued from previous page}.}
    \label{Tab12}
    \end{threeparttable}
\end{table}

    \begin{table}[!h]
    \centering
    \begin{threeparttable}
    \begin{tabularx}{\textwidth}{XXX}
    
    Bob has the possibility to cancel the transaction within the deadline (T1 + F days) after receiving the package and after the transaction is successful. Bob needs to return the package to Alice to receive the refund for his X tokens
    & Possibility for Bob, within the T1 + F days deadline, to cancel the transaction (after [9] and [9'] where Alice informs the package number, after [10] where Bob confirms or does not claim delivery in the T1 + B days deadline, after [12] where Bob is not eligible to receive Y tokens, after [13] where Bob does not respond to the satisfaction questions in the T2 + C days deadline, after [14] where Bob responds that he is satisfied with the satisfaction questions, and after [15] where Alice and Bob have come to an agreement on the relevant conflict in the T3 + D days deadline) once he has received the package and the transaction is considered complete and successful (\textit{right of return} and \textit{return for non-conformity} applicable). In this instance, the transaction is cancelled, the package returned to Alice, the escrow is released and Bob is refunded with his X tokens.
    & \makecell{[C5]} \\ 
    \cmidrule(l r ){1-3}
    
    \end{tabularx}
    \caption{\textit{Continued from previous page}.}
    \label{Tab13}
    \end{threeparttable}
\end{table}


\clearpage
\subsection{Technical Features \& Specifications (Lyzis Trust System)}\label{techfeat&spe}

In this section, we present the Lyzis trust system.

The trust system provides mechanisms and features in a decentralized environment without TTP that significantly increase the level of trust perceived by buyers and sellers towards the Lyzis Marketplace, and thus significantly decrease the perceived risks. This mainly leads to a higher transactional interest and a higher gain of entrants in the Lyzis ecosystem.

\subsubsection{Mechanisms and Incentives to Drive Operational Success}

As stated in Section \ref{feat}, we assume that favorable behaviors must be incentivized to reduce inefficiencies.  Different incentives and mechanisms are thus implemented from the beginning within the native Lyzis environment in order to support and lead to an operational success, including:

\paragraph{Lyzis' own Reputation System.}\label{reputation}
In an online system, reputation\footnote{As defined in \cite{supporttrust}, \textit{Reputation} is represented as "\textit{an expectation about the behavior of an agent based on information about its past behavior}".} allows to express the beliefs or opinions of an individual towards another person or item \cite{trustrepmodel}. In most cases, reputation systems are build to determine the trustworthiness of users and to provide incentives for users to make a fair contribution to the peer-to-peer network \cite{challengesdemarket}.
A reputation management system and its quantification are necessary and thus required within the Lyzis environment, because online exchanges of objects, services or other involve interactions between parties who do not know each other, have no mechanism to trust each other and might actually have no way to  find each other if necessary, as for a physical shop.
The basic idea behind this system is to allow parties to rate each other following an exchange. When a party is considered for an interaction, past evaluations can be aggregated into a score allowing to decide whether to trust them or not. 
In other words, reputation system tries to evaluate the transactions performed between the peers and associate a reputation value to each one. 
Reputation values are then used as selection criteria among each peers. 
The Lyzis reputation system is primarily based on the following elements:

\begin{itemize}
\item[--] Each initial user initially has $50 \%$ of  points at his disposal when creating his account.

\vspace{-0.1cm}

\item[--] After each interaction and as mentioned in the operating algorithm \ref{Fig3}, each party (i.e., the buyer or the seller) have - depending on whether or not the transaction was successful - the possibility to attribute points of good behavior ($+x pts$) or bad behavior ($- y pts$) to the opposite party (these points are  resulting from the "\textit{satisfaction response}" function).

\vspace{-0.1cm}

\item[--] After each interaction and as mentioned in the operating algorithm \ref{Fig3}, each party (i.e., the buyer or the seller) have - depending on whether or not the transaction was successful - the possibility to give a positive comment or a negative comment to the opposite party (these elements are resulting from the "\textit{satisfaction response}" function).
\end{itemize}

The main target of this implementation (i.e., an in-house reputation system) remains to ensure positive interactions and satisfactory operations within Lyzis' decentralized environment.

\paragraph{Economic and Decision-Making Incentive.}\label{ecoincentive}
Since their inception \cite{nakamoto2008bitcoin}, blockchain systems have relied above all on a system of incentives and subsequent rewards, in order to drive  participants to adopt honest behavior within the network.
With the idea of rewarding beneficial behaviors and penalizing those that may be detrimental to the Lyzis environment -- as discussed in the \textit{"participation"} part of Section \ref{feat} --, we introduce  LZSP tokens, dedicated to several different payoff cases, and allowing users to gain economic and decision-making power, i.e., freely transferable against LZS and allowing to vote in the project's DAO (see Section \ref{DAO} and \cite{Lyzistokenomics}).

\paragraph{Staking as Activation Function.}\label{stakingact}
A crypto-currency unit is said to be staked if its holder delegates/renounces the right of relative exchange for a pre-specified period of time \cite{staking}. Hence, a crypto-currency holder who decides to stake is given a monetary reward (in expectation) to compensate for the constraint of not being able to trade their assets for the given period.
Within the Lyzis Marketplace, a seller will have to stake a certain amount of the native LZS token (defined as \textit{Seller Deposit} within the environment) in order to ensure, during his first interactions, his reliability by activating his account. In other words, only seller accounts with an amount staked for a definite duration, will be able to sell items and interact on the platform. 
 The \textit{Seller Deposit} will be effective regardless of the wallet used, i.e., custodial and non-custodial (see Section \ref{custodandno}).

\paragraph{Minimum Deposit Value.} To ensure content with minimal interest in the marketplace and drive optimal value, only assets with a fixed sale value $\geq$ USD 20 shall be depositable.

\subsection{Game Theory Application}\label{gametheory}

The mechanisms described before, whose aim is to foster a good functioning by the presence of incentives to good behaviors and disincentives to bad ones, are supported by a mathematical field called \textit{Game Theory}, which models strategic choices and interactions between different actors. \textit{Game Theory} provides mathematical tools to analyze interactions between rational decision-makers \cite{commnet}.
In a game, each decision maker, as a player, chooses his strategy to maximize his utility given the strategies of the other players.
\textit{Game Theory} can be used and applied within Lyzis Marketplace to analyze the possible strategies that can be employed by users and the interactions between them.
By analyzing the application of this theory, platform users can learn and predict the rational behaviors of others and have optimal reaction strategies based on the notion of equilibrium \cite{gametheory}. Furthermore, \textit{Game Theory} can be used to develop/represent incentive mechanisms that discourage users from performing dishonest behavior or launching potential attacks - as in this case. 

In order to describe or model a situation representing a game (i.e., a decisional interaction situation) four elements must be specified and taken into consideration: \textit{(i)} the set of players (or decision makers); \textit{(ii)} the possible actions for each player; \textit{(iii)} the rules of the game specifying in particular the order in which the players play and when the game ends and \textit{(iv)} the possible outcome of the game for each player and its implication in terms of "\textit{payoff function}" (to do this we need to know the players' preferences).
\\
Considering these elements, we  briefly formalize the \textit{strategies} that can be adopted within the Lyzis Marketplace and the associated results (\textit{payoffs}) in the form of a simultaneous and non-cooperative\footnote{Term widely used in \textit{Game Theory} to express a situation in which we assume that the players are completely independent of their decisions at the time they make their selections.} game with 2 players where:

\begin{center}
    $i_{1}$ = The buyer buying an item from the seller $i_{2}$
    
    \vspace{+0.2cm}
    
    $i_{2}$ = The seller selling an item to the buyer $i_{1}$
\end{center}
 
Also, the sequential nature of this game means that $i_{1}$ and $i_{2}$ decide at the same time on the \textit{strategy} they are going to adopt (when placing the order and validating the sale). 
\\
The possible \textit{strategies} adopted by the players $i_{1}$ and $i_{2}$ are then represented by:

\begin{center}

    $s_{1}$ = Player behaves honestly
    
    \vspace{+0.2cm}
    
    $s_{2}$ = Player behaves dishonestly
\end{center}

Moreover, in this game, each player gets a \textit{payoff} that depends on its own strategy and the strategy of its interlocutor. 

We therefore have the parameters and variables defined where:

\begin{itemize}

    \item \textit{$v_{1}$} = Refers to the monetary value involved in the transaction
    
    \item \textit{$v_{2}$} = Refers to the tangible value relative to the physical asset involved in the transaction
    
    \item \textit{$vo_{1}$} = Refers to a gain of the asset's value - (we assume that, based on the outcome of the transaction, both parties (i.e., buyer and seller) will gain the asset's value, whether they are in possession of the tokens, $v_{1}$, or the asset worth the defined value, $v_{2}$)
    
    \item \textit{$vo_{2}$} = Refers to a loss of the asset's value - (we assume that, based on the outcome of the transaction, both parties (i.e., buyer and seller) will lose the asset's value, whether they are in possession of the tokens, $v_{1}$, or the asset worth the defined value, $v_{2}$)

    \item \textit{$\alpha\sim LZSP_{1}$} = Refers to a gain of Lyzis participation/governance tokens (LZSP)(see \cite{Lyzistokenomics} for more details on distribution mechanisms)

    \item \textit{$\alpha\sim LZSP_{2}$} = Refers to a loss (and/or non-gain) of Lyzis participation/governance tokens (LZSP)(see \cite{Lyzistokenomics} for more details on distribution mechanisms)

    \item \textit{$\beta\sim S_{1}$} = Refers to a maintenance - by the seller - of the LZS tokens initially stacked to validate the operation of his account (see Section \ref{stakingact} for more details) and the possibility for him to withdraw the tokens following the exchange with the returns granted based on the stacking duration (under pre-defined conditions)
    
    \item \textit{$\beta\sim S_{2}$ =} Refers to a loss - by the seller - of the LZS tokens initially stacked to validate the operation of his account (see Section \ref{stakingact} for more details) and the impossibility for him to withdraw the tokens

    \item \textit{$\beta\sim S_{3}$ =} Refers to a loss (non-gain) - by the buyer - of the LZS tokens initially stacked by the seller to validate the functioning of the account (see Section \ref{stakingact} for more details)

    \item \textit{$\beta\sim S_{4}$ =} Refers to a gain - by the buyer - of the LZS tokens initially stacked by the seller to validate the functioning of the account (see Section \ref{stakingact} for more details)
    
    \item \textit{$\delta\sim R_{1}$} = Refers to a positive change in a user's reputation level on the Lyzis Marketplace (see Section \ref{reputation} for more details) - (in parentheses in Tab.\ref{Tab11} as it is not mandatory and depends on the goodwill of the user concerned - but we assume here that agents will do it because they obtain LZSP tokens as a result)
    
    \item \textit{$\delta\sim R_{2}$} = Refers to a negative change in a user's reputation level on the Lyzis Marketplace (see Section \ref{reputation} for more details) - (in parentheses in Tab.\ref{Tab11} as it is not necessarily mandatory and depends on the goodwill of the user concerned - but we assume here that agents will do it because they obtain LZSP tokens as a result)
\end{itemize}

Considering the variables and functions mentioned above as well as the \textit{different strategies}, the \textit{number of players}, and the \textit{associated payoffs}, we obtain Tab.\ref{Tab14} modeling the game theory applied within Lyzis' marketplace.
We assume that no communication is made between the seller and the buyer regarding their adopted strategies, i.e., honest ($s_{1}$) or dishonest behavior ($s_{2}$). We also assume for simplicity that no differences are made between small and large sellers, as well as a non-differentiation between aggressive and conservative strategies.

\vspace{+0.3cm}

\begin{table}[]
\renewcommand{\arraystretch}{2.5}
\setlength{\tabcolsep}{0.12cm}
\begin{center}
\begin{tabular}{|c:c:c|}

\hline
$i_{1}/i_{2}$  & $i_{2}\sim s1$                                                                                                                                            & $i_{2}\sim s2$                                                                                                                                                            \\

\hdashline[3pt/2pt]
$i_{1}\sim s1$ & \begin{tabular}[c]{@{}c@{}}$vo_{1},\alpha\sim LZSP_{1}, \beta\sim S_{3}, (\delta\sim R_{1}); $\\$ vo_{1}, \alpha\sim LZSP_{1}, \beta\sim S_{1}, (\delta\sim R_{1})$\end{tabular} & \begin{tabular}[c]{@{}c@{}}$vo_{1}, \alpha\sim LZSP_{2}, \beta\sim S_{4};$\\ $  vo_{2}, \alpha\sim LZSP_{2}, \beta\sim S_{2}, (\delta\sim R_{2})$\end{tabular} \\

\hdashline[3pt/2pt]
$i_{1}\sim s2$ & \begin{tabular}[c]{@{}c@{}}$vo_{2}, \alpha\sim LZSP_{2}, \beta\sim S_{3}, (\delta\sim R_{2});$\\ $vo_{1}, \alpha\sim LZSP_{2}, \beta\sim S_{1}$\end{tabular}  & \begin{tabular}[c]{@{}c@{}}$vo_{1}, \alpha\sim LZSP_{2}, \beta\sim S_{3};$ \\ $ vo_{1}, \alpha\sim LZSP_{2}, \beta\sim S_{2}$\end{tabular}                 \\ \hline

\end{tabular}
\caption{Payoff matrix based on strategic adoptions.}
\label{Tab14}
\end{center}
\end{table}

\begin{description}
    \item Here are the \textit{strategic possibilities} arising from Tab.\ref{Tab14}.
    
    \item [The case $i_{1} \sim s_{1} / i_{2} \sim s_{1}$ - (Social Optimum/Nash Equilibrium):] The buyer and seller opt for an honest strategy - this is the best outcome possible that favours the well-being of each (seller and buyer) and their personal interests. In this case the sum of the individual utilities is the greatest. We are here in the case of a \textit{social optimum} in the sense of Pareto, maximizing the utility of the relationship for each of the parties. This case also represents the \textit{Nash equilibrium} reached within the relationship between the seller and the buyer, i.e., it is the optimal strategy that each party must adopt in order to maximize its interests and its personal well-being without even knowing or taking into account the opposing strategy.
    
    \item[The case $i_{1} \sim s_{2} / i_{2} \sim s_{1}$:] The seller acts honestly but the buyer decides to adopt a dishonest behaviour - the seller is then paid as if the transaction had taken place correctly (without the LZSPs so he loses nothing), while the buyer is penalized within the network considerably.
    
    \item[The case $i_{1} \sim s_{1} / i_{2} \sim s_{2}$:] The buyer acts honestly but the seller decides to adopt a dishonest behaviour - the buyer is then paid as if the transaction had taken place correctly (without the LZSPs so he loses nothing), while the seller is penalized within the network considerably.
    
    \item[The case $i_{1} \sim s_{2} / i_{2} \sim s_{2}$:] Both parties decide to adopt a dishonest strategy and are then penalized within the network in a considerable way but still less than in the cases: $i_{1} \sim s_{2} / i_{2} \sim s_{1}$ and $i_{1} \sim s_{1} / i_{2} \sim s_{2}$ for the fraudulent party (i.e., the asset and/or the value in tokens are then sent/refunded to the issuing party as much as possible).
\end{description}

\subsection{Minimal Viable Product (\textit{MVP})}\label{mvp}

The \textit{MVP} is a presentable product, beyond prototype, but with limited
features \cite{mvp}. The main target is to prove the functioning of the Lyzis Marketplace to a pre-defined category of actors and then to grow by learning from this first introduction.

\vspace{-0.1cm}

\begin{itemize}
    \item[--] Why initially target \textit{game-related categories}?
\end{itemize}

\vspace{-0.1cm}

Our ultimate goal is to address and accommodate all actors, buyers/consumers and sellers, present online today within the global virtual landscape and to be able to assist them in their daily commercial interactions by bringing a new and innovative model of trust, security, efficiency and transparency and by becoming a reference as a full featured marketplace in the emerging web3 world.
Our platform has been designed for all types of users. Note that  several initiatives have emerged recently  in the field of blockchain technology/web3, but most of them only relate to buying or selling NFTs (e.g., Opensea, KnownOrigin) - no solution exists yet for trading physical assets by leveraging the decentralization principles and this is what Lyzis' ambition is all about.
Consequently, the size of the end market addressed by Lyzis is truly vast and significant. We then decide to initially target a niche of actors, as a first step towards a gradual/scalable expansion process.
Also, in view of the various recent trends\footnote{Globally, the blockchain-based games industry grew by 2,000\% in 2021 and blockchain-based games attracted 1.22 million unique active wallets (UAW) in March 2022. Furthermore, we can also consider the fact - as an indicator and information - that the share of the "Computer \& consumer electronics" category within the retail e-commerce in the US (the most concerned country after China), represented 42.7\% in 2021 and the sales concerning this category are expected to reach \$190.31 billion by the end of 2022. Additionally, the "Computer \& consumer electronics" category generated the highest number of sales on Amazon in 2021 with a total of \$96.03 billion - which definitely illustrates the high level of interest expressed towards this category.}, the Lyzis Marketplace will target the community of gamers\footnote{\textit{Gamers} are defined as users who regularly participate in virtual games, crypto (GameFi) or not, and who we assume have \textit{(i)} like most web users, already made online purchases and \textit{(ii)} express material needs related to the regular use of these games.} present in the NFTs space and the web3 as a whole, allowing us to reach an ideal intersection between classic web users and users used to interacting with digital assets and related substructures (e.g., wallets, chain interactions). The focus will be implemented with the creation of categories within the Lyzis Marketplace that are solely related to this domain\footnote{Mainly ``Video games and consoles / High-Tech networks and hardware / Phones, tablets, computers, connected objects and accessories''.} as well as essentially the establishment of \textit{(i)} several partnerships with GameFi (Gaming and Decentralized Finance) platforms and \textit{(ii)} targeted marketing campaigns through exchanges\footnote{Users of exchanges (CEX/DEX) in the blockchain technology space tend to be familiar with business transactions involving the use of digital assets and related infrastructure}.

\vspace{-0.1cm}

\begin{itemize}
    \item[--] Why using \textit{Polygon Chain}?
\end{itemize}

\vspace{-0.1cm}

Several factors make Polygon (MATIC) stand out as our preferred choice as a layer-2 scaling solution to initially support the Lyzis ecosystem, which includes our decentralized open marketplace for object exchange and e-commerce operations.
Formerly known as the Matic network, the India-based blockchain platform came up with a way to surmount the main challenges faced by the main network for decentralized apps (dApps): Ethereum. Polygon seeks to make Ethereum expand in size and efficiency. As many developers willing to build on Ethereum because it serves their application needs were scared away by the network clogging issues and the high transaction fees, Polygon came to the rescue. 
With our planned marketplace for the exchange of physical objects, we are banking on the decentralization advantage of blockchain technology to reinforce the old trade by barter system for objects to be exchanged between two entities in a P2P way. As we provide participants with a permissionless, inclusive, and interactive environment where a transaction can be carried out between parties without any intermediary involved, we anticipate a successful ecosystem that would cross borders and reach millions of users. 
The innovative solutions that we are proposing to the problems facing centralized traditional marketplaces across borders include low transaction costs. It comes with the urgent need for an adequate level of trust and security to ensure the smooth operation of transactions, and connections as we facilitate exchanges between consumers and producers. 
We depend on Polygon - with its foundation on Ethereum and its added layer that unclogs the Ethereum network for a more efficient process - to provide us with security, trust, privacy, low transaction cost, and data integrity and ensure the best possible launch.

\vspace{-0.1cm}

\begin{itemize}
    \item[--] Why using \textit{USDC} as stablecoin?
\end{itemize}

\vspace{-0.1cm}

It should also be pointed out that USDC will be the preferred stablecoin to use within the Lyzis Marketplace during the MVP phase. This ERC-20 token is one of the most popular crypto assets, the second largest stablecoin in terms of market capitalization at the time of writing and fourth among all crypto-currencies. 
USDC is stable and less volatile because it maintains its price at a 1:1 ratio to the U.S. dollar and is backed by reserves across existing financial institutions. 
Choosing USDC would preserve value for our marketplace users, even over a long period of time and also facilitate trading to more conventional currencies.

\subsection{Custodial and Non-Custodial Wallets}\label{custodandno}

In order to primarily foster access and adoption of digital assets by mainstream ecommerce users entering the Lyzis network, we propose  the implementation of a stamp zone allowing users new to the web3 field to create/generate an internal wallet custodial to Lyzis.
This custodial wallet will then be accessible from the main Dashboard of the project and is intended to be widely used within the marketplace and more globally within the Lyzis environment and its features. Other experienced users will also be able to import non-custodial wallets (i.e., Metamask, Trustwallet) to interact in the Lyzis ecosystem.

\newpage 

\subsection{Further Platform Implementations}\label{future}

Lyzis Marketplace is not only dedicated to the exchange of physical assets and can therefore be opened to a wide range of other possibilities and use-cases.

\subsubsection{Extension by Introducing New Categories}

As a decentralized marketplace, Lyzis Marketplace aims at a maximum and consequent market opening including a multitude of assets of all types. 
Several models and mechanisms are possible regarding the expansion and addition of new categories within the marketplace due to the initial availability of game-related assets (see Section \ref{mvp})  to ensure continuous interest, including the submission of DAO proposals (i.e., \textit{low/medium level} - see Section \ref{DAO}) and their voting.

\subsubsection{Services and Dematerialized Goods}

Considering the functioning of the smart-contract and the platform for an object exchange, it is also possible on the same model but with some operating nuances for users within the Lyzis Marketplace to offer services (e.g., freelancers) or dematerialized goods (e.g., e-books, courses).

Fig.\ref{Fig5} illustrates the representation of a service provider - in this case an engineering/development service - and a customer connection on the Lyzis Marketplace, in which both parties, the provider and the customer, are connected in a decentralized way (i.e. without relying on any TTP).

\vspace{+0.4cm}

\begin{figure}[h!]
\includegraphics[width=0.9\textwidth]{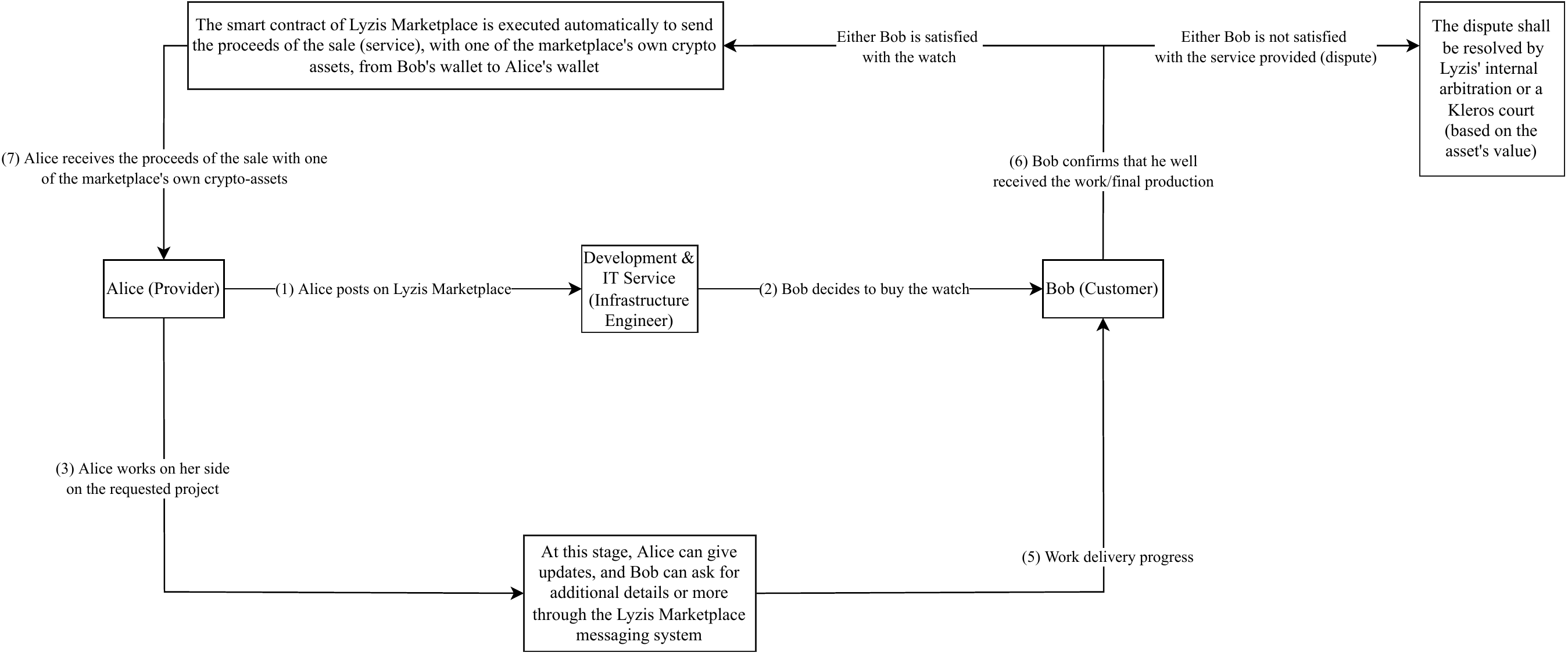}
\center
\caption{How a P2P exchange on the Lyzis Marketplace works in terms of providing a service and/or a dematerialized good.}
\label{Fig5}
\end{figure}

As an illustration, Fig.\ref{Fig6} shows the sequence diagram observed within Lyzis Marketplace, which details the possible interactions among the actors involved in providing a service.

\begin{figure}[H]
\centering
\includegraphics[width=0.85\textwidth]{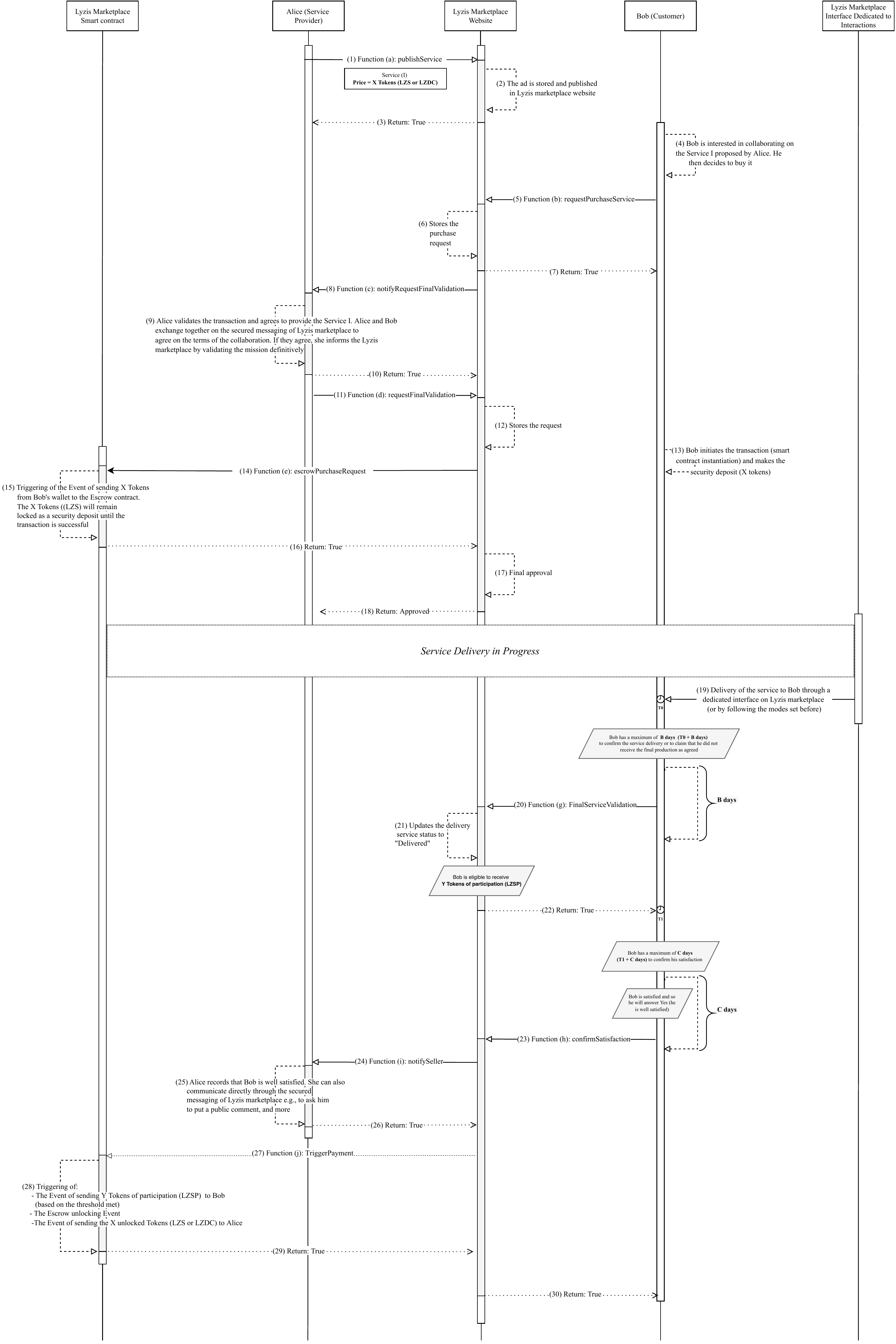}
\caption{UML diagram of a sample workflow of the proposed marketplace (service related).}
\label{Fig6}
\end{figure}

\vspace{-0.25cm}

Finally, Fig.\ref{Fig7} shows a description of the functioning of the algorithm integrated in the Lyzis Marketplace for the provision of a service. Here, we initially assume that there are no applicable "\textit{right(s) of return}" and "\textit{return for non-conformity}" cases.

\begin{figure}[H]
\centering
\includegraphics[width=0.58\textwidth]{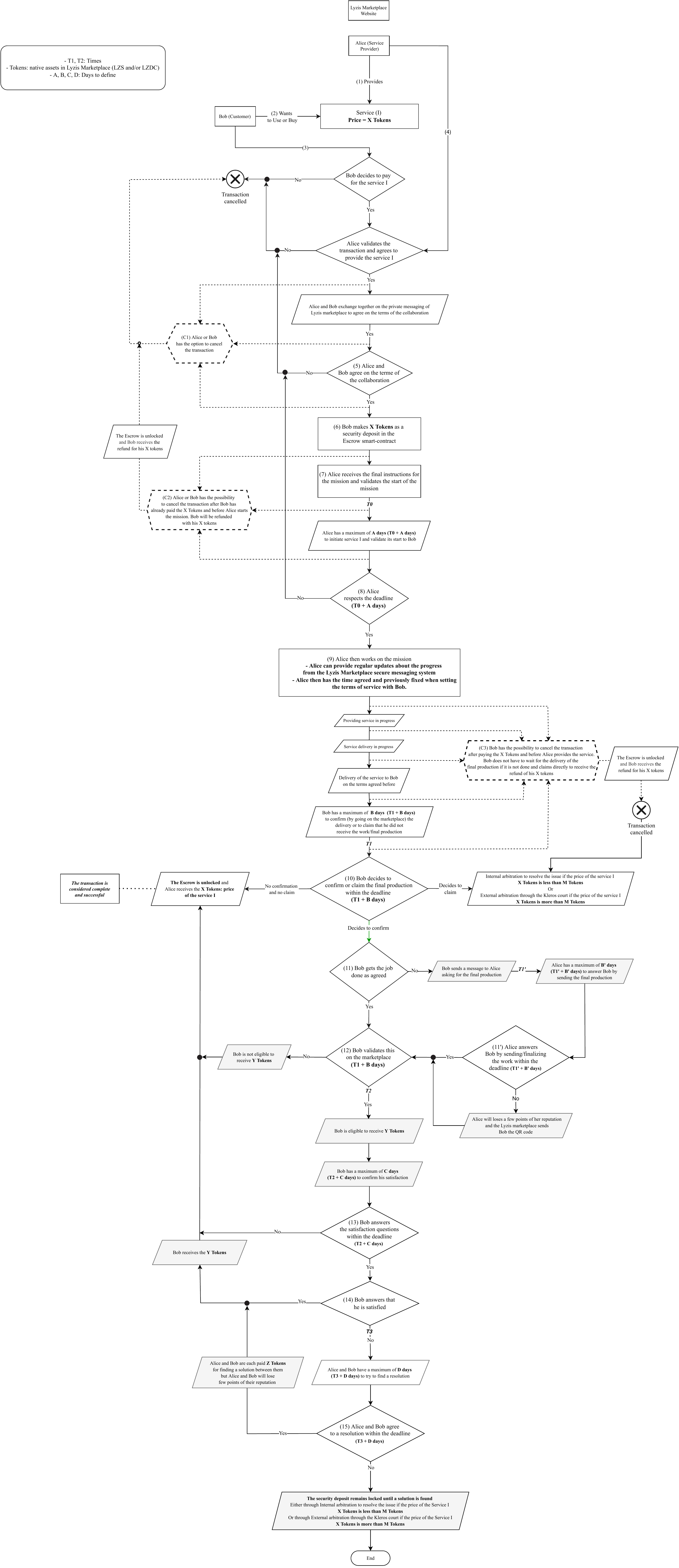}
\captionlistentry{}
\label{Fig7}
\end{figure}
\begin{center}
Figure 6: Technical description of the Lyzis Marketplace algorithm representing step by step the realization of a service provision.
\end{center}

\section{Conflict Management}\label{conflict}

The decentralized nature of the project and consequently of exchanges operating and flowing through Lyzis Marketplace might make it difficult to solve conflicts that arise during interactions implying two peers (seller and buyer).
Decentralized dispute resolution protocols exist, but they mainly incur arbitration costs that are not always necessary and desired by the seller and/or the buyer.
We therefore propose two very distinct conflict resolution systems, depending mainly on the plaintiff's choice and the value of the asset (total value of the transaction).

\subsection{Internal and External Arbitration}

To manage disputes that may arise during interactions on the Lyzis Marketplace, we initially propose to internally process disputes whose value is $\leq$ 50 USD (see Section \ref{thresholds}).
A user engaged within an exchange whose value is less than or equal to the given threshold can claim and report a case directly from the page dedicated to the exchange by using the "\textit{Claim}" button and by filling out a form explaining his request and the related issue.

For disputes where the value is above the given threshold or for complex arbitrations, we suggest using the Kleros protocol as an external arbitration.

\subsection{Kleros Protocol in Lyzis Marketplace \textit{(External Arbitration)}}\label{klerosexplication}

To settle Lyzis Marketplace related cases exceeding the specified threshold\footnote{This measure is driven by arbitration costs that may arise and dependency on arbitration timeframes (see Section \ref{thresholds}).}, we rely on the use of the Kleros justice protocol \cite{lesaege2019short, lesaege2021long}.
Kleros is a blockchain dispute resolution layer that provides fast, secure and cost-effective arbitration for all online disputes (an opt-in court system).
To explain how Kleros can be used in Lyzis Marketplace, let's take the following example of buying/selling an item (or refer to Fig.\ref{Fig8}): 

\begin{itemize}
    \item[--] The seller Alice posted on Lyzis Marketplace a used branded watch to sell with one of the marketplace's own crypto-assets (see Section \ref{assetslyzis}).
    
    \vspace{-0.1cm}
    
    \item[--] Alice has described the characteristics of this watch.
    
    \vspace{-0.1cm}

    \item[--] According to the description of the watch posted by Alice on Lyzis Marketplace, the buyer Bob finds this watch very nice, not very expensive and then he decides to buy it. 
    
    \vspace{-0.1cm}
    
    \item[--] On the one hand, in the ideal case without any dispute, Alice has to send to Bob, in a package format, the watch that corresponds exactly to the description.  Bob has to confirm, after receiving the watch, that the description is correct and that he is happy with the purchase of this watch. Alice will then receive the price of this watch automatically thanks to the smart contracts of Lyzis Marketplace (see Fig.\ref{Fig2}).

    \vspace{-0.1cm}
    
    \item[--] On the other hand, let's assume that in this decentralized marketplace, a dispute has arisen between Alice and Bob because of one or more of the reasons described in Section \ref{litigesreasons}. So, Bob is frustrated, he tries to contact Alice to solve the problem between them but Alice does not answer Bob anymore. In order to resolve this dispute in this decentralized marketplace, Bob will choose to use the Kleros solution by tapping the button "Send to Kleros" and filling out a form explaining his request and problem. 
\end{itemize}

According to the different use cases and graphs provided by the Kleros protocol \cite{lesaege2021long}, we are able to represent in Fig.\ref{Fig8} a classical case of arbitration through the use of external arbitration within the Lyzis Marketplace when a dispute occurs.

\begin{figure}[!h]
\centering
\includegraphics[width=0.6\textwidth]{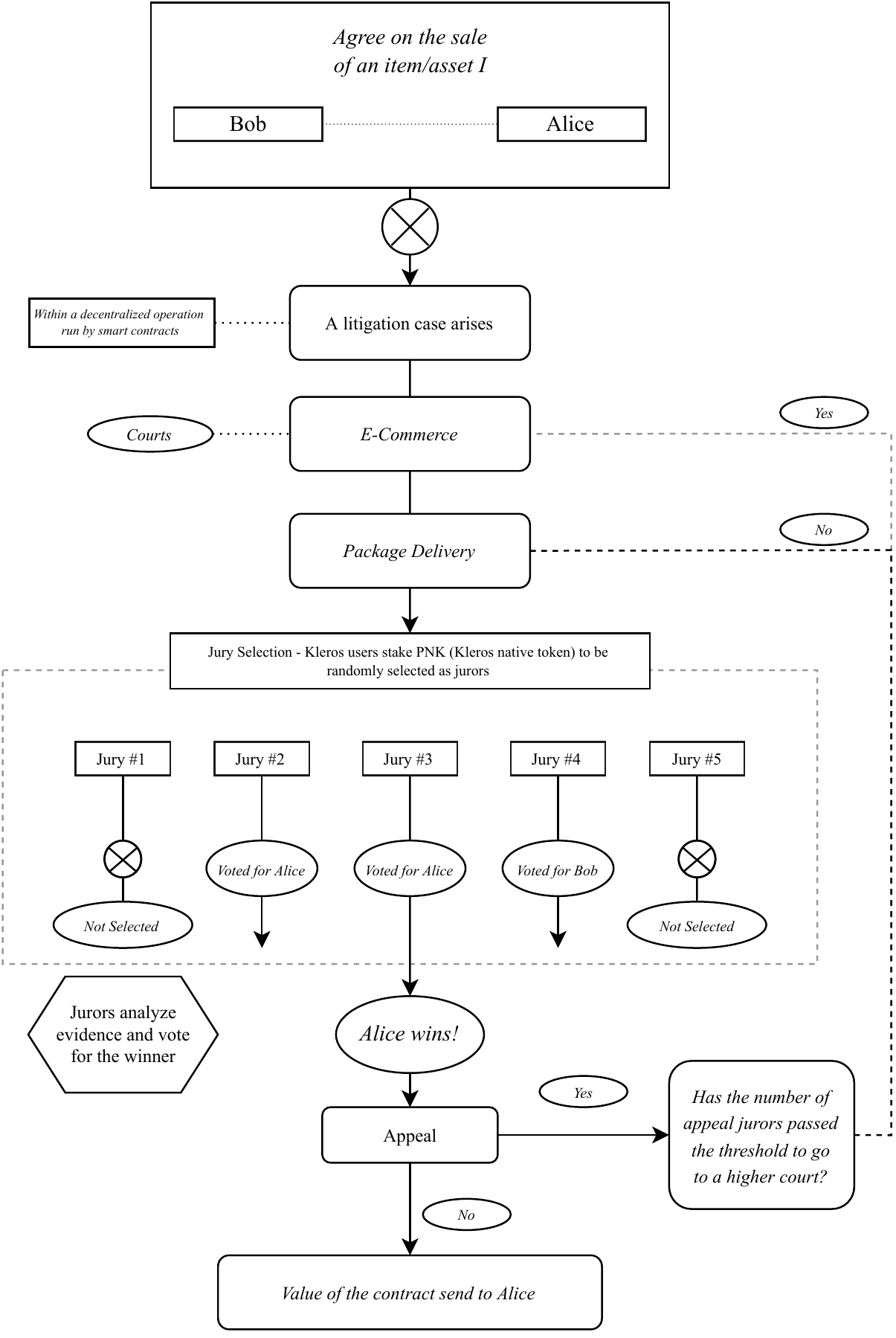}
\captionlistentry{}
\label{Fig8}
\end{figure}
\begin{center}
Figure 7: Description of how a potential dispute is managed within the Lyzis Marketplace when using the Kleros protocol as an external arbitration.
\end{center}

\subsubsection{Steps Involved in Resolving a Dispute}

The conflict resolution process with the use of Kleros protocol is done in different steps.

\begin{enumerate}

    \item A dispute arises in any of the situations discussed in Section \ref{litigesreasons} between Bob and Alice.
    
    \vspace{-0.1cm}
    
    \item The plaintiff, in this case Bob, chose to press the "send/solicit to Kleros" button available on the exchange page on the Lyzis Marketplace. He has to insert the amount of the Kleros fees (jury remuneration) in the smart-contract dedicated to the resolution of the exchange.
    
\vspace{-0.1cm}

    \item The parties (especially/generally the plaintiff (Bob)) are given an evidence period to provide evidence and arguments for their respective cases. This evidence is consistent with the \textit{ERC 1497} standard \cite{ERC1497} defining primarily the requirements for organizing evidence and how it triggers smart contract events \cite{lesaege2021long}.
    
    \vspace{-0.1cm}
    
    \item Candidate jurors express their availability within the Kleros protocol and are then drawn as jurors by self-selecting into a specific court by staking the native Kleros token in that court. The final selection of the jurors is done in a totally random way.
    
    \vspace{-0.1cm}

    \item The randomly selected Kleros jurors are then able to evaluate the evidence that has been submitted, and receive instructions from the dedicated sub-court (in our case - Blockchain > Technical) on how they should reason on the basis of this evidence \cite{klerosdoccourts}.
    
    \vspace{-0.1cm}
    
    \item Each juror then submits a vote\footnote{He essentially submits a hash (vote/salt/address). In Ethereum, the hash used is named \textit{keccak256}.} that reflects the decision made that they believe is the best outcome based on the evidence provided. Various jurors who do not reveal their votes are penalized and other mechanisms are set up to ensure optimal vote aggregation consistency \cite{lesaege2021long}.
    
    \vspace{-0.1cm}
    
    \item The result is revealed to Bob and Alice about their dispute at the end of the period dedicated to the vote of the Kleros jurors. Jurors are encouraged to reveal their votes at the right time, as a penalty may be applied for mistimed disclosure.
    
\end{enumerate}

\subsubsection{Possible Results after Solving a Dispute}

There are obviously two outcomes to consider based on the decision of a Kleros arbitration tribunal to manage a dispute: \textit{(i)} the parties (Bob and Alice) accept the result or \textit{(ii)} one of them believes that the result is unfair.

In the \textit{first case (i)}, the exchange is simply considered completed and the funds (object values in the initial smart contract dedicated to the exchange) are released to the winning party of the conflict (Bob or Alice).

In the \textit{second case (ii)}, it is indeed possible for one of the two parties to "appeal" to the Kleros court if the result in the first instance is considered unfair - so the case is considered by the jury another time. In the event of an appeal, the number of jury members is double that of the previous arbitration plus one\footnote{As the number of juries increases, the relative fees costs will increase in a rather exponential way (to be considered) \textit{(appeal fees = new amount jurors $\sim$ average fee per juror)}\cite{lesaege2021long}. This measure mainly acts as a disincentive for too many appeals and the generation of successive arbitration processes.}.

\subsection{Sources of Disputes between a Seller and a Buyer}\label{litigesreasons}

Different cases of disputes can then arise between the buyer and the seller of an \textit{item I} and this during several phases of the exchange, i.e., pre-sending, per-sending, post-sending. 

Here are then the main reasons\footnote{\textit{Main} because may evolve over time based on observations of exchange operations and potential returned cases of litigation.} to consider:

\vspace{-0.1cm}

    \begin{itemize}
    
    \item The description of the item characteristics is wrong;
    
    \vspace{-0.1cm}
    
    \item One of the two parties involved (seller/buyer) no longer wishes to finalize the exchange;
    
    \vspace{-0.1cm}
    
    \item One of the two parties involved (seller/buyer) does not give any more news to his interlocutor;
    
    \vspace{-0.1cm}
    
    \item The item has not been delivered;
    
    \vspace{-0.1cm}
    
    \item The item received is not the right one and/or the package received is empty;
    
    \vspace{-0.1cm}
    
    \item The item is defective or does not work.

    \end{itemize}
    \vspace{-0.1cm}

In the vast majority of these different litigation cases, a precise description and the provision of as much information as possible concerning the exchange carried out or initiated is always  highly recommended for the complaining party during the evidence phases.

\subsection{On Thresholds and Basket Values}\label{thresholds}

Online stores offer several advantages with respect to brick-and-mortar shops, but it also suffers from additional costs for  order fulfilment and delivery. Many online sellers therefore seek to recoup delivery costs by charging customers for shipping \cite{chen2018shipping}. Different strategies exist, including  contingent free shipping, where the seller
charges a shipping fee only for orders below a minimum order value. The choice of this threshold is known to have an impact on customer behaviour, including the decision to purchase, or not, or to group different items to surpass it, which translates into an impact on the distribution of the values of each transaction. Here, in addition to a possible  threshold for shipping, depending on the sellers' strategy, we also incorporate a threshold for dispute management. The choice of this threshold will be motivated by an optimal balance between its deterring effect on customer engagement, but also the costs associated to the internal versus external arbitration. It is important to note, here, that the distribution of values for a transaction is often a right-skewed distribution, whose mean may significantly  differ from its median. As an example, even in the case of a simple exponential distribution \begin{equation}
p(x) = \lambda e^{- \lambda x}
\end{equation}
where $x$ is the value of a transaction, the mean is $\frac{1}{\lambda}$ while the median is $\frac{\ln 2}{\lambda} \approx \frac{0.67}{\lambda}$, thus much more to the left. This implies that more than half of the transactions have a value below the mean, $1-e^{-1} \approx 63\%$ to be precise. This situation can become even more unbalanced in the case of ''fat-tailed" distributions, like power-laws, often encountered in socio-economic systems \cite{anderson2006long}. 

\section{Economic Overview}

\subsection{Framework}

Many parameters and variables allow us to consider and establish the token economy (increasingly known as "Tokenomics") of Lyzis Labs, closely correlating with the central points that are the structural baselines shared by a wide range of successful Decentralized Finance (DeFi) initiatives.
Consequently, the tokenomics genesis is framed by merging the analysis of a set of positive signs with a design metric of an economic mechanism; this mechanism seeks efficiency and balance in a utility token model while considering multiple key factors, including value creation, velocity control, and economic incentives.

\subsection{Lyzis' digital assets}\label{assetslyzis}

In line with the importance of establishing a strong and viable economy, we envision within the Lyzis environment the implementation of multiple digital assets all dedicated to address the points discussed in Section \ref{feat}, i.e., \textit{utility, participation} and \textit{stability}.

\subsubsection{Lyzis Basic Token - LZS}\label{LZS}

The utility token (LZS) is a digital token issued on the blockchain technology that essentially provide \textit{(i)} a future use of the platform (i.e., security deposit required to activate an account, various staking functions, possibly pay with items and then pay fees generated by SCs instantiations), \textit{(ii)} an access to the Lyzis ecosystem features and various related use cases, and \textit{(iii)} the project's participation and its funding model.

This token derives its value primarily from this right of access and its intended use following the migration of the network used for the MVP to the main network, which stems directly from the demand for its use and exchange in the ecosystem. The LZS does not confer ownership rights to Lyzis Marketplace or the firm's assets, but guarantees future exploitation and project representation through its usage rights, usage expectations, potential price increase and tradability level. The LZS can be exchanged on the platform to guarantee the payment of objects, but is not used as a stable exchange medium in any way.

The distribution mechanisms specific to the digital assets issued by Lyzis Labs are also important variables to consider and play a key role in determining potential future success. We therefore plan to issue the LZS according to a genesis distribution, i.e., a distribution defining the starting conditions\footnote{According to a precise pre-defined allocation allowing to support Lyzis Labs and its development and to be able to plan the phases of the issuance of LZS, e.g., by prioritizing a certain number of tokens linked later to the mainnet.} of the network.

\subsubsection{Lyzis Participation Token - LZSP}\label{LZSP}

The participatory token (LZSP) is a digital token issued on blockchain technology allowing its users to access the various participation modes implemented within the ecosystem. 
Being both  a means of accessing the DAO and other participatory features as well as an asset used to incentivize the entire network, the LZSP is a basic governance token with consequent voting rights in the project's DAO (with a \textit{1 token = 1 vote} ratio), and hence derives its value from the increase in its demand correlated to the various participation needs, the various decisions to be made and the tasks available, the increase in the number of users within the network, and consequently represents a significant incentive to hold and maintain stability, especially with the bridge to freely transfer it, in some instances, with the LZS. For LZS and LZSP, the greater the number of users, the greater the intrinsic value of these assets is likely to be, in a parallel and correlated way (with a significant increase in LPs).

Regarding the LZSP distribution mechanism, a genesis distribution will also be employed and will be effective as soon as the participatory features are implemented in the Lyzis ecosystem, with a specific part allocated to a distribution system categorized as an incentive within the environment. This allows to create a subtle correlation between the number of users participating in the network and the number of users basically only involved in Lyzis Marketplace exchanges. As such, each user will be credited with LZSPs for each participatory behavior leading to a better economic stability, efficiency, activity and global operation, within the Lyzis Marketplace.

\subsubsection{Lyzis Stablecoin - LZDC}\label{LZDC}

LZS is not a short-term store of value and does not serve as a stable intermediary in exchanges. Consequently, the introduction of the LZDC stablecoin with the same usage rights as the above-mentioned LZS in the marketplace, but whose stability ensures equal value transfer when buying/selling items and general exchange within the Lyzis ecosystem is planned. 

In order to ensure adequate stability, we plan to use Paxos \cite{paxos} as a third party providing the trust, technology and financial infrastructure required to implement and maintain the stability of the LZDC. The LZDC will therefore follow a basic USD collateralization (i.e., [1:1]) and will be issued in accordance with the project external roadmap.

\subsection{Relative Work}

A work related to the native economy designed and set up by Lyzis Labs, consisting essentially in studying several parameters/standards to drive and support the project's viability and economic structuring, ensuring a consequent techno-economic base is accompanying this paper \cite{Lyzistokenomics}.

\section{Governance Model}\label{DAO}

The principle of "Decentralized Autonomous Organization" (DAO) was introduced by Daniel Larimer \cite{larimer}. 
In addition, Vitalik Buterin, Ethereum founder, proposed in 2014 that when a DAO was launched, it might be organized to operate without human interactions, provided that the smart contracts were supported by a Turing-complete platform \cite{dao}. 
The DAO concept extends much further than the simple concept of organization, and allows the integration and involvement of a community even more at the center of a given project.
This allows users to have decision-making power and to vote accordingly the network rules, propose initiatives and all this in a transparent, immutable and decentralized way.

We decide to use Lyzis DAO to extend the project participation scope to users of multiple classes, with the main intention to overcome issues of traditional centralized platforms where one entity can dictate and implement rules that users can't/no longer accept without any intervention means.
Governance within the Lyzis environment is decentralized and has three distinct layers of involvement/governance. This allows network users to make changes (minor and major), propose actions, and consequently have decision-making power in the project using the LZSP token.

\begin{enumerate}
    \item \textit{Governance Layer $\sim$ Basic Users}: A user who is present on the Lyzis network but only participates in exchanges/interactions on the marketplace and is therefore not involved in governance (no rights).
    
    \item \textit{Governance Layer $\sim$ Member Users}: A \textit{basic user} who has earned or purchased LZSPs and participates in the various governance procedures of the project through acquired decision-making power (i.e., submission of proposal(s), active participation in voting procedures). A \textit{member} must meet the membership requirements\footnote{The membership requirements are then mainly characterized by \textit{(i)} having a proposal accepted by the DAO, \textit{(ii)} having a certain threshold of LZSP and \textit{(iii)} having a certain level of reputation maintained. Any \textit{basic user} with LZSP can submit proposals at certain levels. Once a proposal is accepted, it becomes a \textit{member user} provided that the requirements are continuously met. \textit{Member users} can propose any type of proposal (minor or major changes) and only these can vote.} in order to govern or risk being converted to a \textit{basic user}.
    
    \item \textit{Governance Layer $\sim$ Delegate Users}:  A \textit{member user} who decides to delegate the decision-making power granted by his LZSPs in possession, by stacking his tokens in the governance contract, to another \textit{member user} with a certain level of reputation within the trust system. This might typically be a vendor user who has received LZSPs as a result of repeatedly engaging in honest and beneficial behavior (see Section \ref{gametheory}) and who wishes to delegate his voting rights to another user because he does not wish to be involved in the governance of the project for a number of reasons, but still wishes to have an impact on the project at his own scale (i.e., because he trusts the so-called "\textit{delegatees}" user).
\end{enumerate}

2. and 3. of the governance layers shall go through the on-chain governance. 
This form of governance has the advantage of being easy to implement in a decentralized and secure manner, and will require costs (fees) for users wishing to be involved (mainly \textit{member users}), acting to some extent as a regulator of the number of submitted proposals. Indeed, low resiliency of off-chain governance as raw governance on the chain leads us to not consider this solution at this time.

\subsection{Proposal Types: \textit{Low/Medium Level} and \textit{High Level}}

For concreteness, two levels of proposal types are defined - \textit{low/medium} and \textit{high}. 
Tab.\ref{Tab15} summarizes these different levels.
In this instance, users must submit their proposals in the desired level (i.e., \textit{low/medium} and \textit{high level}).
If they fail to correctly categorize their proposals, users - regardless of the governance level to which they belong - may be subject to a potential penalty from the committee (i.e., losing a portion of involved tokens or a specific blacklisting level based on action frequency).

\begin{table}[]
\begin{tabular}{|c|c|c|}
\hline
Low and Medium Level Proposals                                                                                                                                                                                                                     & High-level proposals                                                                                                                                   & Time in Proposal Pool \\ \hline
\begin{tabular}[c]{@{}c@{}}Change in the Lyzis product and \\ identity (i.e., with mainly included \\ without limitations changes in \\ the logo, main designs (UI/UX), \\ social networks, Lyzis products, \\ website and features)\end{tabular} & //                                                                                                                                                     & 1 week                \\ \hline
//                                                                                                                                                                                                                                                 & \begin{tabular}[c]{@{}c@{}}Any change in the internal \\ monetary system of Lyzis \\ and related to protocol \\ and internal operations\end{tabular} & 1 month               \\ \hline
\end{tabular}
\caption{Description of the different types of proposals possible within the \textit{Lyzis DAO}.}
\label{Tab15}
\end{table}

\subsection{Voting token-based DAO}

As discussed in Section \ref{LZSP}, the Lyzis DAO is accessed by owning a classic ERC20 governance/participation token, the LZSP \cite{Lyzistokenomics}.
This token stands for voting power in the Lyzis system and takes a basic standard of \textit{1 token = 1 vote} (conventional ratio).
Consequently, a buying power risk may be incurred due to the coordinated purchase of governance tokens by some actors to control decisions and governance at a certain level and hence favor their own personal interests.
To limit this risk, several systems can be set up with especially no further distinction once a given LZSP level is reached and users are \textit{member users} (i.e., beyond a specific threshold, the number of tokens held is no longer considered, the user stays in the maximum governance layer and only has $x$ of tokens available to vote).
Hypothetically, membership in a defined governance layer becomes more important than the number of tokens held within the network at a certain level.
We can also add to this structure the inclusion of a dedicated approval committee.

\subsection{Multi-sig Wallet \textit{(Committee)}}\label{multisig}

A committee meeting dedicated to the validation (or not) of the different high-level proposals, network changes, most of them important, is obviously essential to ensure a good global functioning and to favor the driving of the Lyzis environment towards the best possible operation.
Since the DAO's genesis and based on a three-year cycle, a committee will be elected consisting of five members. 
Each member shall have a private signature which, combined, allows the desired decision to be approved, as long as a quorum of at least 80\% is reached with a 50\% minimum agreement.
This type of structure is made possible by the use of a multi-signature (\textit{Multi-Sig}) wallet linked to each member. Using this form of digital signature, multiple users (i.e., the committee) can approve decisions as a group by combining each of the unique signatures.

\newpage

\subsubsection{Proposal System}

Several steps are necessary for the successful completion of the vote approval of any level \cite{dao2}:
\\
\newline
\textit{\textbf{Creation}} - Each user can create and submit a proposal in the proposal pool. A typical proposal takes the form of a traditional function and essentially consists of the target\footnote{\textit{The addresses on which a user wishes to call a function.}}, value\footnote{\textit{The amount of Layer 1 tokens the user wishes to send to each respective target.}}, signature\footnote{\textit{The signatures of the functions that the user wants to call on each respective target.}}, data call\footnote{\textit{The representation in bytes of the parameters that a user wishes to pass to each respective function.}} and description\footnote{\textit{A description of what the user wishes to do and/or the proposal submitted (the reason for the transaction)}.}.
\\\vspace{-0.2cm}
\newline
\textit{\textbf{Active period}} - Once a proposal has been created and is active, Lyzis users can start voting with their LZSPs (upvote and downvote) and have the deadlines mentioned in the previous Tab.\ref{Tab15}. Voting can be accessed immediately from the contract or project dashboard (\textit{Lyzis Dashboard}).
\\\vspace{-0.2cm}
\newline
\textit{\textbf{Approval (success)}} - Once the vote is approved (by the user members/delegators or the selected committee) a time period is set up. During this period, regardless of the level of proposal, the committee has veto power to override any decision if necessary.
\\\vspace{-0.2cm}
\newline
\textit{\textbf{Queue and Execute}} - Once the vote is approved, anyone can intervene with the proposal ID as an input parameter to call the "queue" function. Then in the same way with the same ID, anyone can subsequently call the "execute" function. Naturally, the steps fail instantly if the vote has not been previously approved.

\subsubsection{Voting Process}

Fig.\ref{Fig9} summarizes how a voting process works for a \textit{low/medium-level} proposal in \textit{Lyzis DAO}.

\begin{figure}[!h]
\centering
\includegraphics[width=0.8\textwidth]{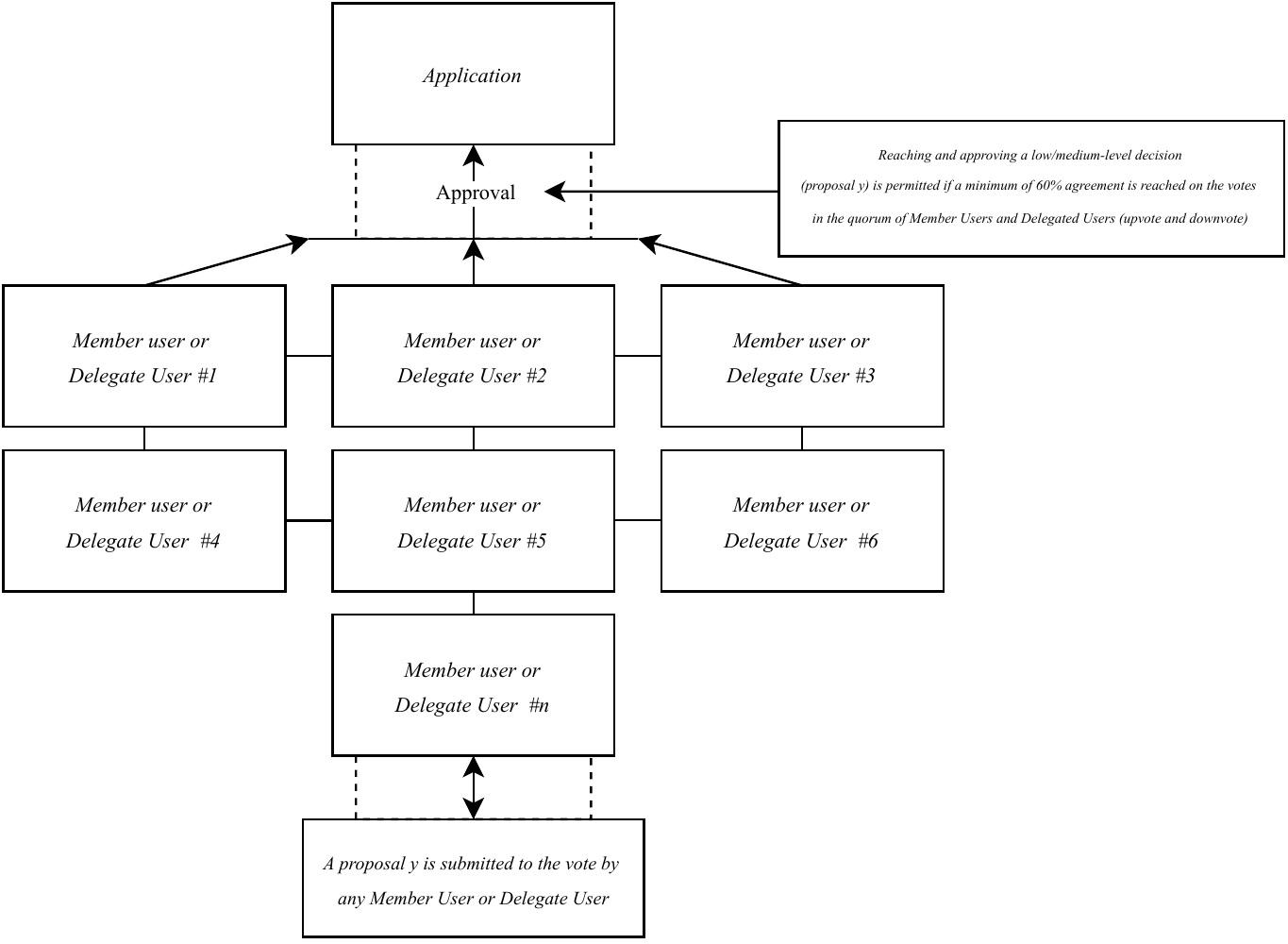}
\captionlistentry{}
\label{Fig9}
\end{figure}
\begin{center}
Figure 8: Description of how submitting a vote for a \textit{low/medium-level} decision works.
\end{center}
\vspace{+0.2cm}

Fig.\ref{Fig10} summarizes how a voting process works for a \textit{high-level} proposal in \textit{Lyzis DAO}.
\vspace{+0.2cm}

\begin{figure}[!h]
\centering
\includegraphics[width=0.9\textwidth]{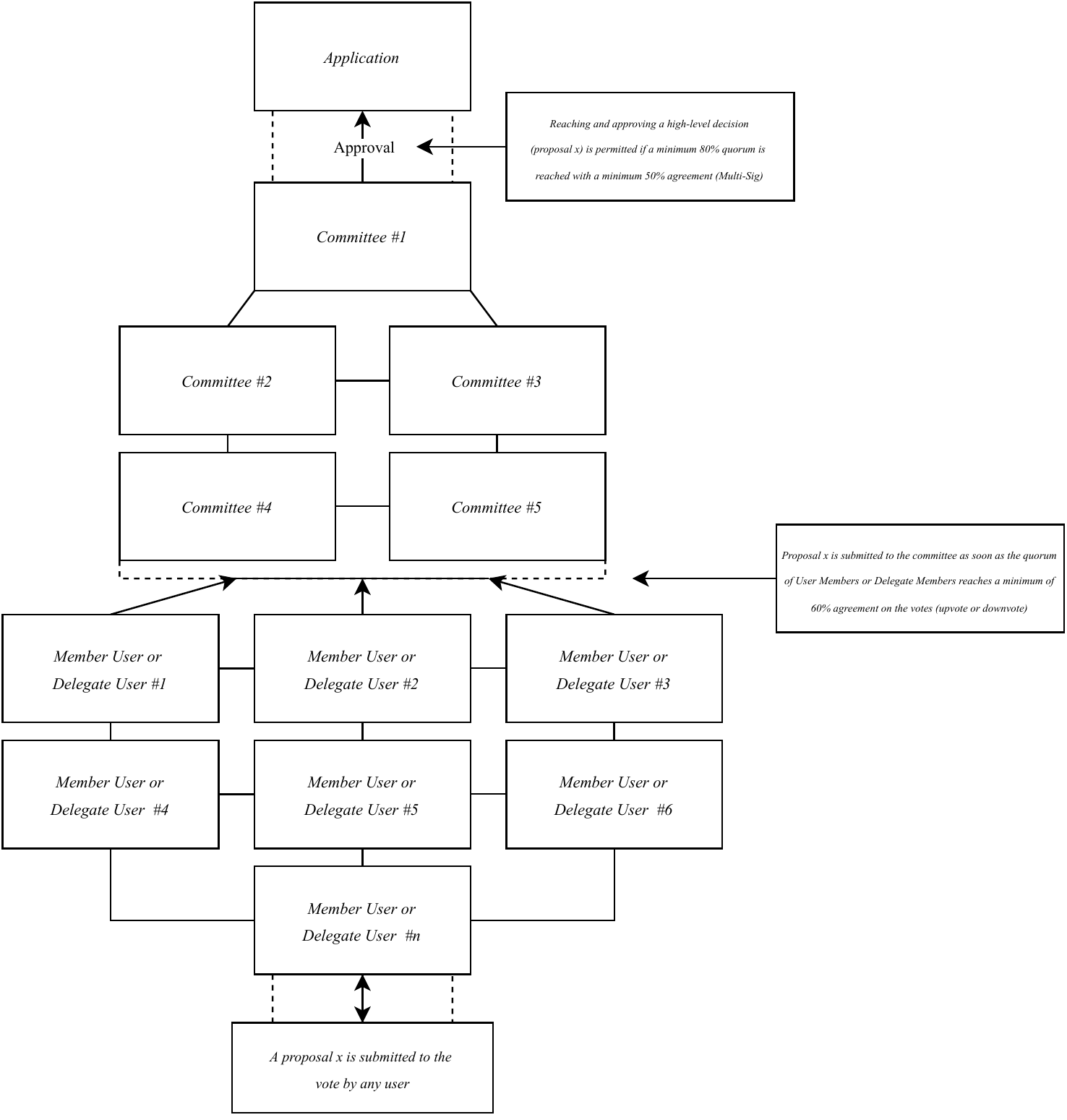}
\captionlistentry{}
\label{Fig10}
\end{figure}
\begin{center}
Figure 9: Description of how submitting a vote for a \textit{high-level} decision works.
\end{center}
\vspace{+0.2cm}

\newpage

\section{Conclusion}

Through \textit{Lyzis Labs}, we  propose a decentralized alternative to classical and centralized e-commerce platforms.
Our solution aims to solve the challenges of centralized platforms on a global scale, by allowing people to exchange (goods for money) without having to rely on trust a central party, in a much more efficient, transparent and safer way that the existing solutions.
 
Rising e-commerce activities and simultaneously the sophistication of digital frauds of all types raise critical concerns related to the trust that users may have towards a platform and drive us to fully consider and support the value of decentralization in a robust way to ensure the efficient operation of a marketplace, \textit{Lyzis Marketplace}.

Our platform relies on several game theoretical mechanisms, providing the necessary incentives/disincentives to stimulate individuals' attitudes and foster beneficial behavior. 
It minimizes the risk that can be perceived within the platform (both from a buying and selling perspective), consequently increasing online trust and removing potential barriers to user activity.
The governing model employed also strongly helps to differentiate from hierarchy-based protocols by involving multi-level users at various management scales and allowing them to be in line with any built-in rules of the system transparently.
Users may then acquire more buying and decision-making power than in conventional centralized marketplaces to profit from the benefits provided by a decentralized system.

\textit{Lyzis Labs} contributes to the decentralized P2P economy by essentially giving to every user equal power, permissionless participation, transparency and common control over data in a structured e-commerce dedicated ecosystem.
\textit{Ultimately, we envision Lyzis to become the exchange standard for physical assets on the blockchain technology.}

\subsection{Future Work}

We have described the main design of the Lyzis technical architecture.
Our current work is focused on the development of the prototype and the smart contracts related to the proposed marketplace, while elaborating the resulting cybersecurity structure and defining the appropriate security assumptions covering the platform, tokens, and governance of the project.

\subsection{Long Term Perspective}

As discussed in this paper, many initiatives related to the use of marketplaces within the web3 linked to NFTs have seen the light of day, very few regarding physical assets.
The challenges outlined in Section \ref{marketchallenges} make it difficult to become a real reference/standard in the field of online commerce on the web3 today, and all the different aspects (e.g. economic, technical, legal environment) of the project must be aligned with the values and needs of its potential actors/users. 
So far, the different projects  related to decentralized marketplaces and dedicated to the exchange of physical objects have shown very weak signs of attraction, low efficiency from the user experience point of view (mainly poor UI/UX design) and therefore an interface too far from the traditional web, thereby complicating the access and many other aspects (see competitor's mapping in Appendix \ref{AppendixA}, Section \ref{AppendixA1}).
The design of Lyzis is built to address these weaknesses, in order to achieve long-term transactional viability.

The idea of a decentralized marketplace being rather basic and fairly generic, its efficient implementation in today's world and its integration into the landscape of tomorrow's web3 world remain challenging and is a key to its success.
This paper has summarised our findings and set the ground for a robust implementation of a decentralized, participatory marketplace.

\subsection{Acknowledgments}

We thank Dr. Abdelhakim Senhaji Hafid (University of Montreal) for his valuable improvements. We thank Dr. Xiaojie Zhu (Abu Dhabi University) and Dr. Natkamon Tovanich (Ecole Polytechique) for their comments.

\addcontentsline{toc}{section}{References}

\bibliographystyle{splncs03}
\bibliography{Mybib.bib}

\begin{thebibliography}{10}

\bibitem{yu2018iotchain}
Bin Yu, Jarod Wright, Surya Nepal, Liming Zhu, Joseph Liu, and Rajiv Ranjan.
\newblock Iotchain: Establishing trust in the internet of things ecosystem
  using blockchain.
\newblock {\em IEEE Cloud Computing}, 5(4):12--23, 2018.

\bibitem{finfraud}
THE EU~CYBERSECURITY AGENCY.
\newblock Financial fraud in the digital space.
\newblock {\em Enisa Europa Eu.}, November 2018.

\bibitem{datastorage}
A.I. Sosin, O.Y. Ivanova, and S.A. Vasilyeva.
\newblock Prospects for implementing blockchain data storage technology as a
  process of digital transformation of society.
\newblock {\em Conference: 2020 International Multi-Conference on Industrial
  Engineering and Modern Technologies (FarEastCon)}, October 2020.

\bibitem{haber1990time}
Stuart Haber and W~Scott Stornetta.
\newblock How to time-stamp a digital document.
\newblock pages 437--455, 1990.

\bibitem{nakamoto2008bitcoin}
Satoshi Nakamoto.
\newblock Bitcoin: A peer-to-peer electronic cash system.
\newblock {\em Decentralized Business Review}, 2008.

\bibitem{zhang2019security}
Rui Zhang, Rui Xue, and Ling Liu.
\newblock Security and privacy on blockchain.
\newblock {\em ACM Computing Surveys (CSUR)}, 52(3):1--34, 2019.

\bibitem{realworldblockchain}
Daniel Hellwig, Arnd Huchzermeier, and Goran Karlic.
\newblock Blockchain in action: Real-world applications.
\newblock {\em In book: Build Your Own Blockchain (pp.173-183)}, May 2020.

\bibitem{dApps}
Gavin Zheng, Longxiang Gao, Liqun Huang, and Jian Guan.
\newblock Decentralized application (dapp).
\newblock {\em In book: Ethereum Smart Contract Development in Solidity}, pages
  253--280, January 2021.

\bibitem{cryptochains}
J.; Felten E.; Miller A.; Goldfeder~S. Narayanan, A.;~Bonneau.
\newblock Bitcoin and cryptocurrency technologies: A comprehensive
  introduction.
\newblock {\em Princeton University Press: Princeton, NJ, USA}, 2016.

\bibitem{TrustworthySmartSystems}
Danda~B. Rawat, Vijay Chaudhary, and Ronald Doku.
\newblock Blockchain technology: Emerging applications and use cases for secure
  and trustworthy smart systems.
\newblock November 2020.

\bibitem{tradbank}
Jeong~Hun Oh and Kevin Nguyen.
\newblock The growing role of cryptocurrency: What does it mean for central
  banks and governments?
\newblock March 2018.

\bibitem{ETHpaper}
Vitalik Buterin.
\newblock Ethereum whitepaper.
\newblock 2015.

\bibitem{paymentResolution}
Aydin Abadi and Steven~J. Murdoch.
\newblock Payment with dispute resolution: Protocol for reimbursing frauds'
  victims.
\newblock {\em https://eprint.iacr.org/}, 2022.

\bibitem{PPcoin}
S.~King and S.~Nadal.
\newblock Ppcoin: Peer-to-peer crypto-currency with proof-of-stake.
\newblock 2012.

\bibitem{Nxt}
Nxt community.
\newblock Nxt whitepaper.
\newblock July 12, 2014.

\bibitem{survey}
Myung-Suk Lee and Kee-Joo Kim.
\newblock Survey on blockchain evolution and proof-of stake consensus
  algorithm.
\newblock {\em International Journal of Engineering Trends and Technology
  69(4):139-141}, April 2021.

\bibitem{permissionandpermissonless}
Andrew Miller.
\newblock Permissioned and permissionless blockchains.
\newblock {\em In book: Blockchain for Distributed Systems Security
  (pp.193-204)}, March 2019.

\bibitem{fincontr}
Alexander~K. Karaivanov.
\newblock Blockchains, collateral and financial contracts.
\newblock {\em Project: Economics of blockchains, Simon Fraser University},
  October 2019.

\bibitem{iotmonit}
Cem Dilmegani.
\newblock Top 17 blockchain applications \& use cases.
\newblock {\em AI Multiple}, April 2020.

\bibitem{votingprocess}
Patricia Baudier, Galina Kondrateva, Chantal Ammi, and Eric Seulliet.
\newblock Peace engineering: The contribution of blockchain systems to the
  e-voting process.
\newblock {\em Technological Forecasting and Social Change}, 2021.

\bibitem{trustissuevote}
David Canon, Guy-Uriel Charles, Edward Foley, and Richard Hasen.
\newblock Restoring trust in the voting process.
\newblock {\em Election Law Journal 20(2)}, May 2021.

\bibitem{ChamolaHealth}
Vinay Chamola, Vikas Hassija, Mohsen Guizani, and Vatsal Gupta.
\newblock A comprehensive review of the covid-19 pandemic and the role of iot,
  drones, ai, blockchain, and 5g in managing its impact.
\newblock {\em IEEE Access PP(99)}, May 2020.

\bibitem{IBM}
Gari Singh and Jonathan Levi.
\newblock Mipasa project and ibm blockchain team on open data platform to
  support covid-19 response.
\newblock 2020.

\bibitem{Dhillon}
Vikram Dhillon, Tailong Xu, and Chirag Parikh.
\newblock Blockchain enabled tracking of physician burnout and stressors during
  the covid-19 pandemic.
\newblock {\em Frontiers in Blockchain 3}, February 2021.

\bibitem{trackmanagehealth}
Kari Korpela, Petr Novotny, Alevtina Dubovitskaya, and Tomi Dahlberg.
\newblock Blockchain design for digital supply chain integration.
\newblock {\em use of blockchain technology in SCM and trade finance - In book:
  Advances in Production Management Systems. Artificial Intelligence for
  Sustainable and Resilient Production Systems (pp.90-98)}, Katarzyna Nowicka.

\bibitem{HealthCare}
Nichola Cooper.
\newblock Editorial: Blockchain in health care.
\newblock {\em Frontiers in Blockchain 4}, February 2022.

\bibitem{decenmarket}
Vishnu~Prasad Ranganthan, Ram Dantu, Aditya Paul, and Paula Mears.
\newblock A decentralized marketplace application on the ethereum blockchain.
\newblock {\em Conference: 2018 IEEE 4th International Conference on
  Collaboration and Internet Computing (CIC)}, October 2018.

\bibitem{Opensea}
Opensea.
\newblock Opensea developer documentation.
\newblock 2017.

\bibitem{challengesdemarket}
Bas van IJzendoorn.
\newblock The challenge of decentralized marketplaces.
\newblock {\em Delft University of Technology, Distributed Systems group},
  March2017.

\bibitem{SybAtt1}
John~R. Douceur.
\newblock The sybil attack.
\newblock {\em Conference: Peer-to-Peer Systems, First International Workshop,
  IPTPS 2002, Cambridge, MA, USA, March 7-8, 2002, Revised Papers}, January
  2002.

\bibitem{SybAtt2}
James Newsome, Elaine Shi, Dawn Song, and Adrian Perrig.
\newblock The sybil attack in sensor networks: Analysis \& defenses.
\newblock {\em Conference: the third international symposium}, April 2004.

\bibitem{SybAtt3}
Roopali Garg and Himika Sharma.
\newblock Prevention techniques for sybil attack.
\newblock {\em INTERNATIONAL JOURNAL OF COMPUTERS \& TECHNOLOGY
  11(10):3060-3064}, December 2012.

\bibitem{adoptionblockchain}
Jiangbo Huangfu, Robert Pinsker, and Hanbing Xing.
\newblock Business strategy and blockchain adoption.
\newblock {\em Conference: Hawaii International Conference on System Sciences},
  January 2022.

\bibitem{statistaecomm1}
Stephanie Chevalier.
\newblock Global retail e-commerce sales 2014-2026.
\newblock {\em Statista, Key Figures of E-Commerce}, September 2022.

\bibitem{statistaecomm2}
Daniela Coppola.
\newblock E-commerce as share of total retail sales worldwide 2015-2021, with
  forecasts to 2026.
\newblock {\em Statista, Key Figures of E-Commerce}, September 2022.

\bibitem{report}
SkyQuest~Technology Consulting.
\newblock Report - global blockchain market, by component, by type, by
  application, by region - forecast and analysis 2022 - 2028.
\newblock {\em Research and Markets}, May 2022.

\bibitem{trustecommerce}
Dan~J. Kim.
\newblock A study of the multilevel and dynamic nature of trust in e-commerce
  from a cross-stage perspective.
\newblock {\em International Journal of Electronic Commerce 19(1):11-64},
  October 2014.

\bibitem{trustdistrust}
Suk-Joo Lee, Cheolhwi Ahn, Kelly~Minjung Song, and Hyunchul Ahn.
\newblock Trust and distrust in e-commerce.
\newblock {\em Sustainability 10(4):1015}, March 2018.

\bibitem{impulsepurchasing}
Ing-Long Wu, Mai-Lun Chiu, and K.-W. Chen.
\newblock Defining key drivers of online impulse purchasing: A perspective of
  both impulse shoppers and system users.
\newblock {\em International Journal of Information Management 36(3):284-296},
  June 2016.

\bibitem{Shopattitude}
Marzieh Zendehdel, Laily Paim, Syuhaily Osman, and Len~Tiu Wright.
\newblock Students' online purchasing behavior in malaysia: Understanding
  online shopping attitude.
\newblock {\em Cogent Business \& Management 2(1):1078428}, December 2015.

\bibitem{drivertrust}
Yassine Jadil, Nripendra Rana, and Yogesh~Kumar Dwivedi.
\newblock Understanding the drivers of online trust and intention to buy on a
  website: An emerging market perspective.
\newblock {\em International Journal of Information Management Data Insights 2
  - 100065}, April 2022.

\bibitem{shopexp}
Weng~Marc Lim.
\newblock Antecedents and consequences of e-shopping: An integrated model.
\newblock {\em Internet Research 25(2):184-217}, September 2015.

\bibitem{hedonic}
Shu-Hao Chang, Wen-Hai Chih, Dah-Kwei Liou, and Yu-Ting Yang.
\newblock The mediation of cognitive attitude for online shopping.
\newblock {\em Information Technology \& People 29(3):618-646}, August 2016.

\bibitem{trustworthiness}
Gomaa Agag and Ahmed~A. El-Masry.
\newblock Why do consumers trust online travel websites? drivers and outcomes
  of consumer trust toward online travel websites.
\newblock {\em Journal of Travel Research 56(3) - Project: Intention
  behaviour}, April 2016.

\bibitem{websiterelated}
Meng-Hsiang Hsu, Li-Wen Chuang, and Cheng-Se Hsu.
\newblock Understanding online shopping intention: The roles of four types of
  trust and their antecedents.
\newblock {\em Internet Research 24(3)}, May 2014.

\bibitem{websitequality}
Mutaz~M. Al-Debei, Mamoun~N. Akroush, and Mohamed~Ibrahiem Ashouri.
\newblock Consumer attitudes towards online shopping: The effects of trust,
  perceived benefits, and perceived web quality.
\newblock {\em Internet Research 25(5):707-733 - Project: Perceived brand
  salience and destination brand loyalty from international tourists'
  perspectives: the case of Dead Sea destination, Jordan}, October 2015.

\bibitem{perceivedusefulness}
Yeolib Kim and Robert~A. Peterson.
\newblock A meta-analysis of online trust relationships in e-commerce.
\newblock {\em Journal of Interactive Marketing 38}, May 2017.

\bibitem{cognitive}
Mary~Ann Eastlick and Sherry Lotz.
\newblock Cognitive and institutional predictors of initial trust toward an
  online retailer.
\newblock {\em International Journal of Retail \& Distribution Management
  39(4):234-255}, March 2011.

\bibitem{Attitude}
Martin Fishbein, Icek Ajzen, and Attitude Belief.
\newblock Belief, attitude, intention, and behavior: An introduction to theory
  and research.
\newblock {\em Contemporary Sociology 6(2)}, MArch 1977.

\bibitem{PurchBehav}
Marzieh Zendehdel, Laily Paim, Syuhaily Osman, and Len~Tiu Wright.
\newblock Students' online purchasing behavior in malaysia: Understanding
  online shopping attitude.
\newblock {\em Cogent Business \& Management 2(1):1078428}, December2015.

\bibitem{Privacy}
Anil Gurung and Manjeri~K Raja.
\newblock Online privacy and security concerns of consumers.
\newblock {\em Information and Computer Security 24(4):348-371}, October2016.

\bibitem{consumbehavgroc}
Phoranee Loketkrawee and Veera Bhatiasevi.
\newblock Elucidating the behavior of consumers toward online grocery shopping:
  The role of shopping orientation.
\newblock {\em Journal of Internet Commerce 17(3):1-28}, October 2018.

\bibitem{plannedbehav}
I.~Ajzen.
\newblock From intentions to actions: A theory of planned behavior.
\newblock January 1985.

\bibitem{privacyimpact}
Chechen Liao, Chuang-Chun Liu, and Kuanchin Chen.
\newblock Examining the impact of privacy, trust and risk perceptions beyond
  monetary transactions: An integrated model.
\newblock {\em Electronic Commerce Research and Applications}, November 2011.

\bibitem{producteval}
Yulia~W. Sullivan and Dan~J. Kim.
\newblock Assessing the effects of consumers' product evaluations and trust on
  repurchase intention in e-commerce environments.
\newblock {\em International Journal of Information Management 39:199-219},
  April 2018.

\bibitem{agriprodcons}
Xiaofei Zhao, Shengliang Deng, and Yi~Zhou.
\newblock The impact of reference effects on online purchase intention of
  agricultural products: The moderating role of consumers? food safety
  consciousness.
\newblock {\em Internet Research 27(2):233-255}, April 2017.

\bibitem{antec}
Charles Dennis, Chanaka Jayawardhena, and Eleni-Konstantina Papamatthaiou.
\newblock Antecedents of internet shopping intentions and the moderating
  effects of substitutability.
\newblock {\em The International Review of Retail Distribution and Consumer
  Research 20(4):411-430}, September 2010.

\bibitem{custominfo}
Hong-Youl Ha and Swinder Janda.
\newblock The effect of customized information on online purchase intentions.
\newblock {\em Internet Research 24(4)}, July 2014.

\bibitem{advertperspect}
Jumin Lee, Do-Hyung Park, and Ingoo Han.
\newblock The different effects of online consumer reviews on consumers'
  purchase intentions depending on trust in online shopping mall: An
  advertising perspective.
\newblock {\em Internet Research 21(2)}, January 2011.

\bibitem{bloggereffects}
Chin-Lung Hsu, Judy Chuan-Chuan Lin, and Hsiu-Sen Chiang.
\newblock The effects of blogger recommendations on customers' online shopping
  intentions.
\newblock {\em Internet Research 23(1):69-88}, January 2013.

\bibitem{onlinegrocershop}
Gary Mortimer, Syed Muhammad~Fazal e~Hasan, Lynda Andrews, and Jillian Martin.
\newblock Online grocery shopping: the impact of shopping frequency on
  perceived risk.
\newblock {\em The International Review of Retail Distribution and Consumer
  Research 26(2):1-22}, January 2016.

\bibitem{ethdao}
Decentralized autonomous organizations (daos).

\bibitem{Lyzistokenomics}
Marwan Zeggari and Renaud Lambiotte.
\newblock Lyzis labs\'\ native economic framework.
\newblock 2022.

\bibitem{trustrelationship}
John Viega and Gary McGraw.
\newblock Building secure software: How to avoid security problems the right
  way.
\newblock January 2001.

\bibitem{supporttrust}
Alfarez Abdul-rahman and Stephen Hailes.
\newblock Supporting trust in virtual communities.
\newblock October 2001.

\bibitem{trustrepmodel}
Y.~Wang and Julita Vassileva.
\newblock Trust and reputation model in peer-to-peer networks.
\newblock {\em Conference: Peer-to-Peer Computing, 2003. (P2P 2003).
  Proceedings. Third International Conference on}, October 2003.

\bibitem{staking}
Cameron Harwick.
\newblock Incentives in blockchain design and applications.
\newblock {\em SSRN Electronic Journal}, January 2020.

\bibitem{commnet}
Zhu Han, Dusit Niyato, Walid Saad, and Tamer Basar.
\newblock Game theory in wireless and communication networks: Theory, models,
  and applications.
\newblock January 2011.

\bibitem{gametheory}
Ziyao Liu, Cong~Luong Nguyen, Wenbo Wang, and Dusit Niyato.
\newblock A survey on applications of game theory in blockchain.
\newblock February 2019.

\bibitem{mvp}
Martin Peacock.
\newblock Minimum viable product -mvp.
\newblock {\em Conference: Internet of Food Things- Project: Biosensors},
  October 2019.

\bibitem{lesaege2019short}
Cl{\'e}ment Lesaege, William George, and Federico Ast.
\newblock Kleros - short paper v1. 0.7.
\newblock September 2019.

\bibitem{lesaege2021long}
Cl{\'e}ment Lesaege, William George, and Federico Ast.
\newblock Kleros - long paper v2. 0.2.
\newblock July 2021.

\bibitem{ERC1497}
Sam Vitello, Cl{\'e}ment Lesaege, and Enrique Piqueras.
\newblock Erc 1497: Evidence standard.

\bibitem{klerosdoccourts}
Court - the heart of the kleros dispute resolution protocol.
\newblock {\em Github - Technical Documentation}, 2021/2022.

\bibitem{chen2018shipping}
Chaoqun Chen and Donald Ngwe.
\newblock {\em Shipping fees and product assortment in online retail}.
\newblock Harvard Business School, 2018.

\bibitem{anderson2006long}
Chris Anderson.
\newblock {\em The long tail: Why the future of business is selling less of
  more}.
\newblock Hachette UK, 2006.

\bibitem{paxos}
Paxos.
\newblock Stablecoin-as-a-service.
\newblock {\em https://paxos.com/stablecoin-as-a-service/}.

\bibitem{larimer}
Daniel Larimer.
\newblock The hidden costs of bitcoin.
\newblock {\em LetsTalkBitcoin/online}, 2021.

\bibitem{dao}
Vitalik Buterin.
\newblock Daos, dacs, das and more: An incomplete terminology guide.
\newblock 2014.

\bibitem{dao2}
Patrick Collins.
\newblock What is a dao? what is the architecture of a dao? (how to build a dao
  - high level).
\newblock February 2021.

\bibitem{statistaecomm3}
Daniela Coppola.
\newblock Global number of digital buyers 2014-2021.
\newblock {\em Statista, Key Figures of E-Commerce}, October 2021.

\bibitem{shopifyreport}
Greg Bernhardt.
\newblock Global ecommerce sales growth report for 2021-2026.
\newblock {\em Shopify Report}, April 2022.

\bibitem{occstrategy}
Mostyn Goodwin, Jan Bergmann, and Alex Birch.
\newblock Trading places online, online marketplaces rise to dominance.
\newblock {\em OC\&C Strategy Consultants}, 2022.

\end{thebibliography}

\newpage

\appendix

\section{Appendix}\label{AppendixA}

\subsection{\textit{Market Details}}\label{AppendixA1}

Strong target market indicators (for informational purposes only):

\begin{itemize}
    \item[-] The global online sales market is worth \$5.5 trillion (2021) - with forecasts to reach \$7.5 trillion by 2025 \cite{statistaecomm1}.
    \item[-] Worldwide retail sales represent 21\% of the global online sales market (2021) - with forecasts to reach 24\% by 2025 \cite{statistaecomm2}.
    \item[-] There are 2.14 billion global digital shoppers (2021) - or about 27.2\% of the world's population that regularly shops online \cite{statistaecomm3}.
    \item[-] Of the total global retail sales in 2022, 20.3\% is expected to come from online purchases \cite{shopifyreport}.
    \item[-] By 2025, online marketplaces will overtake first party eCommerce channels in established categories (as defined below), such as clothing, travel, and books \cite{occstrategy}.
    \item[-] The global blockchain market is valued at \$4,67 billion (2021) and is projected to grow from \$10.02 billion by the end of 2022 to \$163.83 billion by 2029 \cite{report}.
\end{itemize}

Additionally, Tab.\ref{tab16} and \ref{Tab17} summarize the main issues experienced by users within centralized structures (so-called "trust issues") with Ebay and Facebook Marketplace as references.
Finally, Tab.\ref{Tab18} summarizes the main differentiation points of Lyzis with its competitors in the centralized and decentralized online commerce landscape.

\vspace{+0.5cm}

{\bf  Why Ebay and Facebook Marketplace as references?}

Currently, Ebay and Facebook marketplace stand as the most visited online marketplaces in the world and are often used as a reference for central protocols when it comes to selling or buying an item online. On each of the above platforms, it is up to the buyer to ensure the reliability of the seller with whom he wishes to interact and vice-versa, the seller must ensure the reliability of the buyer. 
With the use of blockchain and smart contracts within the Lyzis Marketplace to manage exchanges in a decentralized way, as mentioned throughout this paper, none of the actors have to know whether or not they can and should trust another peer actor before engaging in a transaction.

\vspace{+0.25cm}

{\bf  An adequate local environment to launch a decentralized marketplace?}

The launch of a decentralized marketplace and its successful implementation depends first and foremost on its integration into a favorable environment with mainly \textit{(i)} a high level of internet adoption, \textit{(ii)} established distribution infrastructures and related postal services, \textit{(iii)} a diverse, fragmented buyer base, \textit{(iv)} present users likely to use digital assets as well as \textit{(v)} an overall favorable usage of online commerce.

We note that Lyzis Marketplace also has the potential to trigger the development of online retail in developing countries, where the access to banks and credit cards is limited. 
The exact launch location (during MVP) of the marketplace will be determined in the next phases of the project to ensure an optimal initial adoption.

\begin{table}[htpb]
\resizebox{\textwidth}{!}{%
\begin{tabular}{|l|l|l|l|}
\hline
\textbf{Platform} & \textbf{Issue type} & \textbf{Solution} & \textbf{Part} \\ \hline
\begin{tabular}[c]{@{}l@{}}Ebay \\ (1000+ observed \\ reviews - 18\% good/\\ excellent, 5\% average \\ and 77\% low/poor)\end{tabular} & \begin{tabular}[c]{@{}l@{}}Inefficient/deplorable after-sales \\ service (i.e., ignoring complaints, \\ unacceptable delays, no protection \\ for sellers but only for buyers, \\ defective dispute management, no \\ management when there is no res-\\ ponse from the vendor, no study in \\ dispute - automatic responses,...)\end{tabular} & \begin{tabular}[c]{@{}l@{}}The use of decentralization to limit \\ the number of necessary interventions, \\ the use of a third-party protocol for \\ decentralized dispute management \\ (Kleros) and the implementation of \\ the Lyzis trust system\end{tabular} & 20.6\% \\ \hline
 & \begin{tabular}[c]{@{}l@{}}Cybersecurity (i.e., lack of user con-\\ trol, cancellation/deletion of \\ submitted reviews - unreliable rating \\ system (faked/bought), misleading \\ item descriptions, unjustified penalties \\ applied, identity verification problem \\ and consequent lack of access, bank \\ account number to be provided and \\ direct debits without any autho-\\ rizations,...)\end{tabular} & \begin{tabular}[c]{@{}l@{}}Use of wallets during interactions (no\\ credit card or bank data required), use\\ of decentralization to increase \\ transparency and implementation \\ of several trust mechanisms extracted\\ from the game theory to foster \\ the overall smooth operation \\ (reputation and participation)\end{tabular} & 17.7\% \\ \cline{2-4} 
 & \begin{tabular}[c]{@{}l@{}}Package noted delivered but never \\ received\end{tabular} & \begin{tabular}[c]{@{}l@{}}Automatic tracking and use of a \\ smart contract with the use of QR code\end{tabular} & 12.9\% \\ \cline{2-4} 
 & \begin{tabular}[c]{@{}l@{}}Defective, damaged or non-\\ conforming receipt and/or im-\\ possible return\end{tabular} & \begin{tabular}[c]{@{}l@{}}Automatic management without \\ any TTP intervention\end{tabular} & 15.2\% \\ \cline{2-4} 
 & \begin{tabular}[c]{@{}l@{}}Abuse of guarantees (i.e., either \\ third parties such as PayPal or \\ directly Ebay)\end{tabular} & \begin{tabular}[c]{@{}l@{}}Automatic tracking and use of a \\ smart contract with the use of QR code\end{tabular} & 7.7\% \\ \cline{2-4} 
 & \begin{tabular}[c]{@{}l@{}}Account desactivated without \\ any reason and abusive procedures\end{tabular} & \begin{tabular}[c]{@{}l@{}}Automatic management without any \\ TTP intervention and gouvernance part\end{tabular} & 10.3\% \\ \cline{2-4} 
 & Blocked funds & \begin{tabular}[c]{@{}l@{}}Automatic management without any \\ TTP intervention\end{tabular} & 15.6\% \\ \hline
\end{tabular}%
}
\caption{Main Ebay issues and Lyzis related solutions.}
\label{tab16}
\end{table}

\begin{table}[htpb]
\resizebox{\textwidth}{!}{%
\begin{tabular}{|l|l|l|}
\hline
\textbf{Platform} & \textbf{Issue type} & \textbf{Solution} \\ \hline
Facebook Marketplace & Scams of all kinds and false buyers & \begin{tabular}[c]{@{}l@{}}Initial deposit implemented (staking), \\ KYC initial application, Trust system\end{tabular} \\ \hline
 & \begin{tabular}[c]{@{}l@{}}Possible physical meetings \\ (i.e., assaults, thefts)\end{tabular} & Only online interactions possible \\ \cline{2-3} 
 & \begin{tabular}[c]{@{}l@{}}No after-sales service (e.g., \\ proposal to contact the seller \\ only to settle a dispute)\end{tabular} & \begin{tabular}[c]{@{}l@{}}The use of a third-party protocol for \\ decentralized dispute management \\ (Kleros) and the implementation of \\ the Lyzis trust system\end{tabular} \\ \cline{2-3} 
 & \begin{tabular}[c]{@{}l@{}}Ads not automatically approved \\ (i.e., without valid reasons)\end{tabular} & \begin{tabular}[c]{@{}l@{}}Automatic management without any \\ TTP intervention\end{tabular} \\ \cline{2-3} 
 & \begin{tabular}[c]{@{}l@{}}User blocked and unable to continue \\ interacting on the marketplace\end{tabular} & \begin{tabular}[c]{@{}l@{}}Automatic management without any \\ TTP intervention and gouvernance part\end{tabular} \\ \cline{2-3} 
 & \begin{tabular}[c]{@{}l@{}}Cybersecurity (i.e., lack of user \\ control, cancellation/deletion of \\ submitted reviews - unreliable \\ rating system (faked/bought), mis-\\ leading item descriptions, un-\\ justified penalties applied, identity \\ verification problem since only \\ an account is required,...)\end{tabular} & \begin{tabular}[c]{@{}l@{}}Use of wallets during interactions, use \\ of decentralization to increase transpa-\\ rency and allow everyone to own their \\ data, implementation of several trust \\ mechanisms extracted from the game \\ theory to foster the overall smooth \\ operation (reputation and participation), \\ Automatic management without \\ any TTP intervention\end{tabular} \\ \hline
\end{tabular}%
}
\caption{Main Facebook Marketplace issues and Lyzis related solutions.}
\label{Tab17}
\end{table}

\begin{landscape}
\begin{table}[]
\resizebox{\textwidth}{!}{%
\begin{tabular}{|
>{\columncolor[HTML]{FFFFFF}}l |
>{\columncolor[HTML]{FFFFFF}}l |
>{\columncolor[HTML]{FFFFFF}}l |
>{\columncolor[HTML]{FFFFFF}}l |
>{\columncolor[HTML]{FFFFFF}}l |
>{\columncolor[HTML]{FFFFFF}}l |
>{\columncolor[HTML]{FFFFFF}}l |
>{\columncolor[HTML]{FFFFFF}}l |
>{\columncolor[HTML]{FFFFFF}}l |
>{\columncolor[HTML]{FFFFFF}}l |
>{\columncolor[HTML]{FFFFFF}}l |}
\hline
                                                                                                       & \begin{tabular}[c]{@{}l@{}}Centralized protocols \\ (e.g., Ebay, Facebook\\ Marketplace)\end{tabular} & SaFeX                                                                                                                              & \begin{tabular}[c]{@{}l@{}}Bluebuerry \\ Network\end{tabular}                              & \begin{tabular}[c]{@{}l@{}}Origami \\ Network\end{tabular}                                                              & Mercure                                                                  & \begin{tabular}[c]{@{}l@{}}Darkdot \\ Network\end{tabular}                              & Gamb.io                                                                         & Particl.io                                                                                         & \begin{tabular}[c]{@{}l@{}}Phore \\ Blockchain\end{tabular}                                             & \begin{tabular}[c]{@{}l@{}}Lyzis Protocol\\ \& Marketplace\end{tabular}                                                                                                           \\ \hline
UI/UX                                                                                                  & \textit{High}                                                                                                         & \textit{Low/Poor}                                                                                                                  & \textit{Low/Poor}                                                                          & \textit{Low/Poor}                                                                                                       & \textit{Medium}                                                          & \textit{Medium}                                                                         & \textit{Good}                                                                   & \textit{Good}                                                                                      & \textit{Good}                                                                                           & \textit{High}                                                                                                                                                                     \\ \hline
\begin{tabular}[c]{@{}l@{}}Protocol - \\ Architecture\end{tabular}                                     & \textit{Centralized}                                                                                                  & \textit{Native PoW}                                                                                                                & \textit{API provider}                                                                      & \textit{None}                                                                                                           & \textit{None}                                                            & \textit{PoC}                                                                            & \textit{None}                                                                   & \textit{Native PoS}                                                                                & \textit{Native PoS}                                                                                     & \textit{Native PoS}                                                                                                                                                               \\ \hline
\begin{tabular}[c]{@{}l@{}}Included Trust \\ Mechanisms and \\ Reputation System\end{tabular}          & \textit{\begin{tabular}[c]{@{}l@{}}Yes (only Reputation \\ System)\end{tabular}}                                      & \textit{None}                                                                                                                      & \textit{None}                                                                              & \textit{Yes}                                                                                                            & None                                                                     & \textit{None}                                                                           & Yes                                                                             & None                                                                                               & None                                                                                                    & \textit{\begin{tabular}[c]{@{}l@{}}Yes (High Level \\ Implementation)\end{tabular}}                                                                                               \\ \hline
Governance Aspects                                                                                     & \textit{None}                                                                                                         & \textit{None}                                                                                                                      & \textit{None}                                                                              & \textit{Yes}                                                                                                            & \textit{Yes}                                                             & \textit{Yes}                                                                            & \textit{Yes}                                                                    & \textit{\begin{tabular}[c]{@{}l@{}}None (only \\ decentralized \\ moderation)\end{tabular}}        & \textit{None}                                                                                           & \textit{Yes}                                                                                                                                                                      \\ \hline
Cybersecurity Focus                                                                                    & \textit{Good/High}                                                                                                    & \textit{None}                                                                                                                      & \textit{None}                                                                              & \textit{None}                                                                                                           & \textit{None}                                                            & \textit{None}                                                                           & \textit{None}                                                                   & \textit{Low/Poor}                                                                                  & \textit{Low/Poor}                                                                                       & \textit{High}                                                                                                                                                                     \\ \hline
\begin{tabular}[c]{@{}l@{}}Incentives for Good \\ Operation (monetary \\ system included)\end{tabular} & \textit{None}                                                                                                         & \textit{None}                                                                                                                      & \textit{None}                                                                              & \textit{None}                                                                                                           & \textit{None}                                                            & \textit{None}                                                                           & \textit{\begin{tabular}[c]{@{}l@{}}Medium \\ Implementation\end{tabular}}       & \textit{\begin{tabular}[c]{@{}l@{}}Medium \\ Implementation\end{tabular}}                          & \textit{None}                                                                                           & \textit{High}                                                                                                                                                                     \\ \hline
\begin{tabular}[c]{@{}l@{}}Go-to-Market \\ Strategy\end{tabular}                                       & \textit{//}                                                                                                           & \textit{None}                                                                                                                      & \textit{None}                                                                              & \textit{None}                                                                                                           & \textit{NFT Drop}                                                        & \textit{None}                                                                           & \textit{None}                                                                   & \textit{None}                                                                                      & \textit{None}                                                                                           & \textit{\begin{tabular}[c]{@{}l@{}}Targeting the Gaming \\ Market to MVP and \\ Scalable Expansion\end{tabular}}                                                                  \\ \hline
Asset Used                                                                                             & \textit{Fiat(s)}                                                                                                      & \textit{\begin{tabular}[c]{@{}l@{}}Native Safex \\ Cash\end{tabular}}                                                              & \textit{USDT or DAI}                                                                       & \textit{CB + Swap}                                                                                                      & \textit{Altcoins}                                                        & D4RK/USD                                                                                & \textit{Native GMB}                                                             & \textit{\begin{tabular}[c]{@{}l@{}}Coin integration \\ (e.g., Bitcoin, \\ ETH)\end{tabular}} & \textit{Native PHR}                                                                                     & \textit{Native LZS + LZSP}                                                                                                                                                        \\ \hline
Operation System                                                                                       & \textit{\begin{tabular}[c]{@{}l@{}}Trusted third party \\ control\end{tabular}}                                       & \textit{\begin{tabular}[c]{@{}l@{}}Series of linked \\ transactions recor-\\ ded at regular block \\ intervals (PoW)\end{tabular}} & \textit{\begin{tabular}[c]{@{}l@{}}Integration on \\ e-commerce \\ platforms\end{tabular}} & \textit{\begin{tabular}[c]{@{}l@{}}Professional \\ stores - Inte-\\ gration on \\ e-commerce \\ platforms\end{tabular}} & \textit{\begin{tabular}[c]{@{}l@{}}3 Multi-Sig \\ Contract\end{tabular}} & \textit{\begin{tabular}[c]{@{}l@{}}Escrow - Dual \\ Deposit Toll \\ (DDT)\end{tabular}} & \textit{\begin{tabular}[c]{@{}l@{}}Professional/\\ Private stores\end{tabular}} & \textit{\begin{tabular}[c]{@{}l@{}}2-party Escrow \\ Mechanism\end{tabular}}                       & \textit{\begin{tabular}[c]{@{}l@{}}Multi-sig \\ Transactions + \\ Decentralized \\ Escrow\end{tabular}} & \textit{\begin{tabular}[c]{@{}l@{}}Decentralized escrow \\ system, QR code usage, \\ Internal and External \\ Arbitrations (Kleros), \\ Game Theory Appli-\\ cation\end{tabular}} \\ \hline
\end{tabular}%
}
\caption{Competitive landscape and main differentiation points of Lyzis Labs.}
\label{Tab18}
\end{table}
\end{landscape}

\addcontentsline{toc}{section}{Glossary}
 
\glsaddall
\printglossary[type=main,nonumberlist]

\addcontentsline{toc}{section}{Acronyms}
\printglossary[type=\acronymtype,nonumberlist]

\thispagestyle{empty}
\addcontentsline{toc}{section}{Disclaimer}
\subsubsection*{Disclaimer}

\begin{itemize}
    \item We certify that the information provided in this white paper and on the Lyzis Labs website does not constitute: investment advice, financial advice, trading advice or any other type of advice.
    \item You should not consider the contents of this white paper or the Lyzis Labs website as investment advice, financial advice, trading advice or any other type of advice.
    \item Please conduct your own due diligence  and consult your financial advisor before making any investment decisions.
    \item Lyzis Labs project, including, but not limited to, Lyzis Labs website, smart contracts, assets and all Lyzis applications as presented in this conceptual paper, is not a licensed, unlicensed or exempt financial or payment service of any kind and in any jurisdiction.
    \item Any terminology used in this white paper, on the Lyzis Labs website, or in any Lyzis application is intended to serve as a basic reference only, with no actual or legal meaning of the same terms in a regulated and/or financial environment. Lyzis smart contracts are security audited, permanent and cannot be modified in any way. 
    \item  The Lyzis Token is strictly a "Utility" token in all jurisdictions and is not and cannot be considered a "Security" or regulated token of any kind. 
    \item Lyzis Labs project is in no way an electronic money and/or a fiat currency. This white paper by itself does not constitute a contract or contractual agreement of any kind, nor does it constitute an invitation, solicitation or offer to invest in the Lyzis Labs project or to acquire or use the Lyzis tokens in any manner whatsoever with the expectation of profit. 
    \item Any user of Lyzis represents and warrants that he or she has received appropriate technical, administrative, regulatory and legal advice before and after accessing and/or reading this white paper or the Lyzis Labs website, and/or using any part or element of Lyzis Labs project (including Lyzis Tokens). 
    \item The user hereby acknowledges and agrees that there is a high risk inherent in accessing, acquiring or using any type of blockchain and/or crypto system, token, platform, software or interface, including Lyzis Labs project, and further waives any and all claims of any kind against any member of the community directly or indirectly involved with Lyzis Labs project, for any damages suffered, including total loss. Use is at your own risk. 
\end{itemize}

\newpage

\thispagestyle{empty}
\addcontentsline{toc}{section}{Disclosure}
\subsubsection*{Disclosure}

\begin{itemize}
    \item  The information described in this white paper is preliminary and is subject to change at any time. 
    \item This paper may also contain ``Forward-Looking Statements'' that generally relate to future events or our future performance. ``Forward-Looking Statements'' include, but is not limited to, for the Lyzis Labs project, its anticipated performance, the expected development of its business and projects, the execution of its vision and growth strategy, and the completion of projects underway, in development or under consideration.
    \item Forward-looking statements represent the beliefs and assumptions of our management only as of the date of this presentation. These statements are not guarantees of future performance and undue reliance should not be placed on them.
    \item Such forward-looking statements necessarily involve known and unknown risks, which may cause actual performance and results in future periods to differ materially from any projections expressed or implied herein.
    \item Lyzis Labs project does not undertake to update forward-looking statements. Although forward-looking statements represent our best expectations at the time they are made, there can be no assurance that they will prove to be accurate, as actual results and future events could differ materially. The reader is cautioned not to place undue reliance on forward-looking statements.
\end{itemize}
}    
\end{document}